\documentclass[journal]{IEEEtai}

\usepackage{color,array}

\usepackage{graphicx}

\usepackage{booktabs}
\usepackage{amssymb}
\usepackage[nocompress]{cite}
\usepackage{amsmath}
\usepackage{subfigure}
\usepackage{multirow}
\usepackage{multicol}
\usepackage{stfloats}
\usepackage{float}
\usepackage[ruled,linesnumbered,vlined]{algorithm2e}
\usepackage{algpseudocode}
\usepackage{epstopdf}
\usepackage{changes}
\definechangesauthor[name={Shuer}, color=orange]{S.}
\newtheorem{definition}{Definition}

\begin{document}

\title{PANDORA: Deep graph learning based COVID-19 infection risk level forecasting}

\author{Shuo Yu, \IEEEmembership{Member, IEEE}, Feng Xia, \IEEEmembership{Senior Member, IEEE}, Yueru Wang, Shihao Li, Falih Febrinanto, \\and Madhu Chetty
\thanks{S. Yu is with School of Computer Science and Technology, Dalian University of Technology, Dalian 116024, China. (email: shuo.yu@ieee.org).}
\thanks{Y. Wang is with Department of Mathematics, National Tsing Hua University, Hsinchu, 30013, Taiwan. (e-mail: yueru.wang333888@outlook.com).}
\thanks{S. Li is with School of Software, Dalian University of Technology, Dalian 116620, China. (e-mail: shihao\_leee@outlook.com).}
\thanks{F. Xia, F. Febrinanto and M. Chetty are with Institute of Innovation, Science and Sustainability, Federation University Australia, Ballarat, VIC 3353, Australia. (e-mail: f.xia@ieee.org, falihgozifebrinanto@students.federation.edu.au, madhu.chetty@federation.edu.au).}
\thanks{Corresponding author: Feng Xia.} 
}
\markboth{ IEEE Transactions on Computational Social Systems,~Vol.~14, No.~8, JULY~2022}%
{Yu \MakeLowercase{\textit{et al.}}: PANDORA: Deep graph learning based COVID-19 infection risk level forecasting}
\maketitle

\begin{abstract}
COVID-19 as a global pandemic causes a massive disruption to social stability that threatens human life and the economy. Policymakers and all elements of society must deliver measurable actions based on the pandemic's severity to minimize the detrimental impact of COVID-19. A proper forecasting system is arguably important to provide an early signal of the risk of COVID-19 infection so that the authorities are ready to protect the people from the worst. However, making a good forecasting model for infection risks in different cities or regions is not an easy task, because it has a lot of influential factors that are difficult to be identified manually. To address the current limitations, we propose a deep graph learning model, called PANDORA, to predict the infection risks of COVID-19, by considering all essential factors and integrating them into a geographical network. The framework uses geographical position relations and transportation frequency as higher-order structural properties formulated by higher-order network structures (i.e., network motifs). Moreover, four significant node attributes (i.e., multiple features of a particular area, including climate, medical condition, economy, and human mobility) are also considered. We propose three different aggregators to better aggregate node attributes and structural features, namely, Hadamard, Summation, and Connection. Experimental results over real data show that PANDORA outperforms the baseline method with higher accuracy and faster convergence speed, no matter which aggregator is chosen. We believe that PANDORA using deep graph learning provides a promising approach to get superior performance in infection risk level forecasting and help humans battle the COVID-19 crisis.
\end{abstract}



\section{Introduction}
\label{sec:introduction}

\IEEEPARstart
{T}{he} Coronavirus disease 2019 (COVID-19) is a global pandemic that is presently considered a potential threat to humankind has spread rapidly, causing substantial loss of life and economic crisis worldwide. To minimize the devastating effects of the pandemic, developing the best policies is necessary with considering the conditions of specific regions and the level of severity of the pandemic. One of the best policies that have proven to be effective in reducing the spread of COVID-19 is by implementing social distancing~\cite{Benzell14642,hsiang2020effect,zhang2020changes} or, in an extreme way, by ruling a lockdown in certain areas. However, this kind of policy consequently leads to secular stagnation and inconvenience to people's lives. Locking down the cities and limited human interaction raises new problems as complex as the pandemic itself. The policymakers should consider any aspects and not recklessly generalize the policies to all regions. The most appropriate way to stabilize the conditions is by implementing relatively strict policies to prevent the pandemic from further spreading in the areas with high infection risk~\cite{hsiang2020effect} and not to lock down the regions with lower infection risk; instead, it is better to develop recovery policies to ensure the social and economic system running normally in less affected regions. Therefore, it is crucial to have such a forecasting system to predict and estimate the infection risk of COVID-19 so that the policymakers can get a better insight to produce strict regulations selectively and balance conditions against the pandemic~\cite{Yu2021TCSSDDDM,Liu2020CSDE,walker2020impact, Dehningeabb9789}.
\begin{figure}[t]
     \centering
     \includegraphics[width=.45\textwidth]{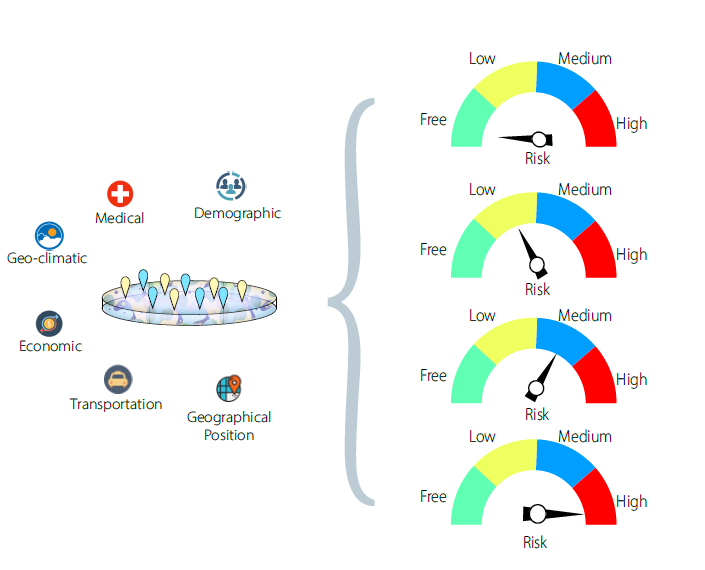}
     \caption{Infection risk level forecasting for certain region.}
     \label{fig:introductiondiagram}
\end{figure}


As investigating COVID-19 infection for a specific region (e.g., city, state, country, area) is very important, other important factors need to consider such as including demographic characteristics (e.g., population size and inhabitants density), medical conditions (e.g., number, capacity, and distribution of hospitals), geo-climatic conditions (e.g., temperature and humidity), etc. Apart from the epidemiology perspective~\cite{Benzell14642}, there are some other factors have been investigated to understand how COVID-19 spread~\cite{chang2020mobility,wong2020evidence}. The infection mechanism~\cite{Metcalf368,fu2020beaware} has also been studied to minimize the transmission risk. Those kinds of studies are fundamental to enable the awareness of possible unseen factors for the COVID-19 spread. Moreover, understanding the interaction of human mobility with the transportation system is also vital because that interaction might contribute to the spread of the pandemic~\cite{Chinazzi395}. It is imperative to learn that pattern of human mobility to provide insight for a better prevention system in combating COVID-19~\cite{wang2020model}.

The success of deep learning to address many real-world tasks can be used to address those unsolved problems. We understand the significant challenges of this study are all related to the graph data and mostly come from an irregular domain, and then we choose deep graph learning~\cite{Xia2021TAI} instead of regular deep learning methods to understand patterns and structural information between entities from graph data. Deep graph learning has proven as a prominent method of addressing complicated and sophisticated real-world issues for graph data in many fields, including natural language processing, healthcare, image processing, etc. Some studies have been proposed to develop forecasting predictions using deep graph learning, and they have shown the significance performance in forecasting the future patterns~\cite{hou2020network,shen2017multilabel}. By employing higher-order structures, deep graph learning methods show better performance to capture implicit relations and multivariate relations among entities in networks~\cite{chang2018structure}. Moreover, it is important to understand graph motifs, graphlets, and other kinds of network substructures, as they have great benefits for modeling spatial dependencies in transportation networks~\cite{10.1145/3292500.3330869, 9369147, yu2020offer, xu2020multivariate }. 

Apart from some advanced solutions in different tasks using deep graph learning, there are still many challenges in forecasting infection risk for COVID-19:
\begin{itemize}
\item The infection risk is difficult to be quantified, and this problem needs to be formulated first.
\item The infection risk of a specific region might be caused by many factors, which means that the forecasting method should consider comprehensive attributes.
\item Most importantly, human mobility should be considered when forecasting infection risk, even though we are aware of the fact that human mobility is difficult to be calculated for since geographical positions and transportation conditions are significantly different in every region.
\end{itemize}

In general, we use an integrated deep graph learning algorithm to solve the challenges mentioned above. The general diagram of the problem to solve is shown in Fig.~\ref{fig:introductiondiagram}. Medical conditions, economic conditions, and the number of infectious cases are considered as node attributes, the task is to forecast the risk level of certain region. We propose a deep learning method called PANDORA (Deep graph learning-based COVID-19 infection risk level forecasting) to predict the infection risk for specific regions. PANDORA considers comprehensive factors relevant to COVID-19 infection risk, including geographical relations among different areas, transportation frequency, climate, medical condition, economy, and human mobility. A transportation network is built based on geographical location information and the mobility of people. We collected flight data from Transportation Security Administration (TSA) and location data from The United States Census Bureau (USCB) to support the experiment. Considering the transportation routes among different regions, we recognize these routes as structural characteristics in the network. Integrating with other significant factors, including climate features, medical condition, economic level, and human mobility, the proposed model can predict certain regions' COVID-19 infection risk level. In addition, experimental results show that the proposed model PANDORA can identify high-risk areas with higher accuracy than the baseline method. The major contributions of this work can be summarized as follows.

\begin{itemize}
	\item{\textbf{Problem formulation of COVID-19 infection risk level forecasting:}} Even though WHO (World Health Organization) has announced the standard of infection risk level evaluation, measuring and predicting COVID-19 infection risk level is a complex problem of significant importance. However, this problem has not been formally formulated yet. This work aims to predict the COVID-19 infection risk level (i.e., risk-free, low-risk, medium-risk, and high-risk) for diverse regions according to WHO standards. We formulate the model to enable a solution for COVID-19 infection risk level forecasting using those standard levels as label prediction. By doing it, we believe that our model can be generalized because it is in line with many countries' policies.

  \item{\textbf{A deep graph learning approach for COVID-19 infection risk level forecasting:}} As one of the first applications of deep graph learning in COVID-19 data mining, we propose an innovative, integrated deep graph learning framework called PANDORA to predict COVID-19 infection risk of specific regions. PANDORA employs a higher-order network structure (i.e., network motif) to formulate geographical structural features, significantly enhancing network representation learning. By integrating structural features and node attributes, PANDORA outperforms the baseline method in accuracy and convergence speed.
	\item{\textbf{Integrating comprehensive factors in COVID-19 infection risk level forecasting:}} The chance of getting infected with COVID-19 may depend on many factors. However, the impact of these factors hasn't yet been well studied. Herein we explore multiple factors pertinent to COVID-19 infection, including geographical position relationships and transportation frequency among different regions, climate, medical condition, economic status, and human mobility in a particular area. The correlation relationships between these factors and infection risk level are analyzed. At last, multiple factors are integrated as node features to enhance forecasting.
\end{itemize}

The rest of this paper is organized as follows. Section~\ref{sec:rel} introduces related work containing the factors that affect the spread of COVID-19. The start-of-art algorithms in graph learning are also introduced. Section~\ref{sec:pre} provides preliminaries and formally defines the infection risk level forecasting problem. Section~\ref{sec:me} illustrates the details of the proposed method PANDORA. Section~\ref{sec:exp} presents the experimental results, including correlation analytics between COVID-19 infection and different characteristics of regions, comparison results of PANDORA and the baseline method, and results of different aggregators, respectively. Section~\ref{sec:con} concludes the paper.

\section{RELATED WORK}
\label{sec:rel}
\subsection{COVID-19 Data Mining}

How to cope with COVID-19 has become a big challenge for almost all countries worldwide. Many factors affect the spread of respiratory infectious diseases, such as the population size and temperature, medical capacity, and economic development in specific regions. By analyzing the impact of these factors on the pandemic, it does not only make people better understand the pandemic but also predicts the outbreak~\cite{Liu2021BDR,Kissler860}.

Many existing studies have investigated the factors that may affect the spread of COVID-19.
Lolli et al.~\cite{lolli2020impact} studied the role of meteorological conditions on pandemic transmission. It is found that the pandemic transmission prefers dry and cool environmental conditions. By investigating the impact of environmental temperature in the global spread of COVID-19, Huang et al.~\cite{huang2020optimal} discovered that the most suitable temperature for COVID-19 transmission ranges from 5$^{\circ}$C to 15$^{\circ}$C. Laxminarayan et al.~\cite{laxminarayan2020epidemiology} studied the age distribution of the patients. The result shows that the transmission risk is significantly higher in children and young adults. Furthermore, there is fact that mortality also reflects based on several factors such as age and gender. Xiong et al.~\cite{xiong2020mobile} used mobile device location data to investigate the relationship between human mobility and virus transmission. The positive correlation between mobility inflow and the number of infections is discovered by analyzing simultaneous equations. Moghadas et al.~\cite{moghadas2020projecting} illustrated that the growing COVID-19 outbreak could challenge the health centre capacity. More people will be infected if the health care facilities. Hao et al.~\cite{hao2020understanding} used mobility data to simulate the transmission of COVID-19 to investigate the impact of population density and mobility. The results show that the impact of population mobility is far greater than the population density on COVID-19 transmission, so population mobility plays a key role during the transmission of the pandemic. Shuvo et al.~\cite{shuvo2020simulating} employed artificial agent-based simulation to emphasize the importance of social distancing. If social isolation cannot be activated shortly, then constructing hospitals at the initial stage of the pandemic can greatly increase the treatment capacity.

\subsection{Deep Graph Learning}

Graph convolution network (GCN), a method for semi-supervised learning on graph, was firstly proposed by Kpif and Welling~\cite{Kipf2016Semi}. Seo et al.~\cite{seo2018structured} modified the standard GCN and proposed the graph convolutional recurrent network (GCRN) model with the long short-term memory (LSTM) mechanism to  address dynamic networks. Veličković et al.~\cite{velivckovic2017graph} proposed graph attention networks (GATs), which used self-attention~\cite{zhang2019self} to construct the graph attentional layer. The importance of all neighbors of the current node is considered, then it updates the node representation according to the attention of its neighbor node. Gao et al.~\cite{gao2018large} proposed a learnable graph convolutional layer (LGCL) to enable regular convolution operations on generic graphs. Meanwhile, they used the sub-graph training method to make LGCNs suitable for large-scale graphs training. By using the clustering structure of graphs, Chiang et al.~\cite{chiang2019cluster} proposed a Cluster-GCN training algorithm that can train a very deep network. In each step, the node blocks related to the dense subgraph are identified by the graph clustering algorithm~\cite{yin2017local,yu2020detecting}, and limit the neighborhood search within the subgraph. That solution can significantly improve memory and computational efficiency without compromising test accuracy. Schlichtkrull et al.~\cite{schlichtkrull2018modeling} proposed R-GCNs, that was developed specifically to handle the highly multi-relational data characteristic, and proved to be effective in classification tasks. Ying et al.~\cite{ying2018graph} developed a data-efficient GCN algorithm PinSage, which is based on highly efficient random walks to structure the convolutions and design a novel training strategy that relies on harder-and-harder training examples to improve robustness and convergence of the model.

Nowadays, graph convolutional networks have been widely used in many graph-based applications. For instance, Rahimi et al.~\cite{Rahimi2018Semi} proposed a multi-view geolocation model based on a graph convolutional network, which can infer social media user geography location by using text and network information. Girshick et al.~\cite{6909475} proposed RCNN by using adjacency matrices of networks and CNNs, which can identify key nodes in complex networks. Tian et al.~\cite{tian2020ra} proposed a relational aggregation graph convolutional network (RA-GCN) to describe knowledge graphs. RA-GCN can extract more entity and relationship information and optimize entity classification and link prediction. Guo et al.\cite{guo2019attention} modeled the dynamic spatial-temporal correlations with traffic data and employed the attention mechanism into graph convolutions to forecast traffic flows. Specifically, common standard convolutions are applied to describe the temporal features to improve prediction accuracy. Ktena et al.\cite{ktena2017distance} proposed a new metric learning method to evaluate the distance between graphs. At the same time, this method is applied to irregular graphs by spectral graph theory.

To take advantage of graph learning~\cite{Xia2021TAI}, in this work, we propose to apply deep graph learning into COVID-19 infection risk level forecasting~\cite{fu2020beaware,ye2020community}. Network motifs are proven to be effective in network representation learning methods, which are also applied in many downstream tasks. Unlike the trajectory model, network motifs can better formulate multiple relations among several adjacent regions so that structural features can be better described in the network representation learning process. Meanwhile, more node features can be taken into consideration. To address the challenge caused by the complexity of the problem, we devise an innovative deep graph learning framework. Multiple factors have been explored in the design of our approach.

\section{Preliminaries}
\label{sec:pre}

\subsection{Network Motif}
Network motifs are recurrent and statistically significant subgraphs or patterns of a larger graph in comparison to the randomized networks~\cite{yu2019motifs,xia2019survey}. Therefore, the probability that the frequency of a certain motif appears on a randomized network is greater than the probability of a real-world network. The network motif is formally defined based on several constraints, i.e., $P,U,D$. The threshold $P$ is used to qualify whether a certain motif is over-represented in the real-world network or not. $P$ can be estimated by $z$-$score$s, which is defined in Eq.~\eqref{eq:zscore}. The threshold $U$ defines a quantitative minimum to establish significance as shown in Eq.~\eqref{eq:u}. $D$ is the threshold ensuring the minimum difference between $f_{ori}$ and $\bar f_{rand}$, which can be formalized as Eq.~\eqref{eq:d}.

\begin{equation}
z-score(G_k)= \frac{f_{ori}-{\bar f_{rand}}}{std(f_{rand})}
\label{eq:zscore}
\end{equation}


\begin{equation}
f_{ori} \geq U 
\label{eq:u}
\end{equation}

\begin{equation}
f_{ori} - \bar f_{rand} > D \times \bar f_{rand}
\label{eq:d}
\end{equation}
wherein, $G_k$ is a certain network, $f_{ori}$ is the frequency of a given motif in $G_k$, while $f_{rand}$ is the frequency in a randomized network. Moreover, $std(f_{rand})$ represents the standard deviation of motif frequency in different randomized networks. The notations and symbols used in this paper are shown in Table~\ref{tab::notation}.

\begin{table}[htbp]
	\centering
	\caption{Notations and symbols.}
	\resizebox{0.48\textwidth}{!}{
		\begin{tabular}{ccl}
			\toprule
			Classifications & Notations & \multicolumn{1}{c}{Description} \\
			\midrule
			\multirow{6}[2]{*}{Graph }
			& $\mathcal{V}$     & Node set of graph $\mathcal{G}$. \\
			&$\mathcal{E}$     & Edge set of graph $\mathcal{G}$. \\
			& $\mathbf{A} $    & Adjacency matrix of graph $\mathcal{G}$. \\
			&$\mathbf{D}$     & Degree matrix of graph $\mathcal{G}$. \\
			& $\mathbf{x}$   & Graph feature vector.\\
			& $\mathcal{Y}$     & City risk level set. \\
			\midrule
			\multicolumn{1}{c}{\multirow{15}[2]{*}{Parameter\newline{} }} 
			& $\mathbf{SFT}$   & The Structural Feature Tensor. \\
			&$\mathbf{AET}$   & Aggregated Embedding Tensor. \\
			& $\mathbf{AFT}$   & Attribute Feature Tensor. \\
			& $\odot$  & Hadamard product operation. \\
			& $\mathbf{X}$     & Feature matrix. \\
			& $\mathbf{\bar A}$    & Renormalized adjacency matrix. \\
			&  $\mathbf{\bar D}$    & Renormalized degree matrix. \\
			&  $\sigma(\cdot)$  & Activation function. \\
			&   $Agg(\cdot)$ & Aggregation process. \\
			&  $SM(\cdot)$ & Softmax function. \\
			& $\mathbf{H}$ &Feature matrix in AFT or SFT.\\
			&  $\mathbf{K,P}$ & Feature matrix in AFET and SFET, respectively.\\
			&   $\mathbf{R}$ &Feature matrix in AET.\\
			&   $\mathbf{\theta}$   &Weight parameter vector.\\
			&   $\mathbf{W},\Theta$  &Weight parameter matrix to be trained. \\
			\bottomrule
	\end{tabular}}%
	\label{tab::notation}%
\end{table}%

\subsection{Problem Formulation}
Aiming to predict infection risk for certain regions, we first formulate this task by labeling a specific region $Region$ with associated features $Feature=(F_s, F_n)$. $F_s$ refers to structural features, and $F_n=(demo, med, geo-clim, eco)$ refers to node attributes. For a static graph, the problem can be transferred to a node classification task: given the graph with feature at all-time series $F = ({F_s^T},{F_n^T})$, our goal is to determine the risk level at a specific time. For dynamic graph, the problem can be transferred to a time series classification task: given the graphs at all-time series, $\mathcal{G} =\{G_1, G_2, ……, G_n\}$, the object is to predict a class label $c$ for a given length time series $X$ whose label is unknown.

Specifically, risk levels are used as node labels in this work. According to the recommendation of WHO, risk level of certain region is formally defined based on the number of infectors $N_i$ in two weeks:
\begin{itemize}
\item If $N_i=0$, then the corresponding region is defined as \emph{risk-free};
\item If $0<N_i\leq 150$, then the corresponding region is in \emph{low-risk} level;
\item If $150<N_i\leq 750$, then the corresponding region is in \emph{medium-risk} level;
\item If $N_i>750$, then the corresponding region is defined as \emph{high-risk} level.
\end{itemize}

\section{Design of PANDORA}
\label{sec:me}

In this section, the proposed PANDORA is introduced in detail. The motif structure and tensors integration are employed in the design. The proposed model can better reflect the connections between nodes and multivariate relationships. The graph learning tasks can improve accuracy, significantly speed up the convergence, and reduce the time cost in the learning process.
\begin{figure*}[!htb]
     \centering
     \includegraphics[width=1\textwidth]{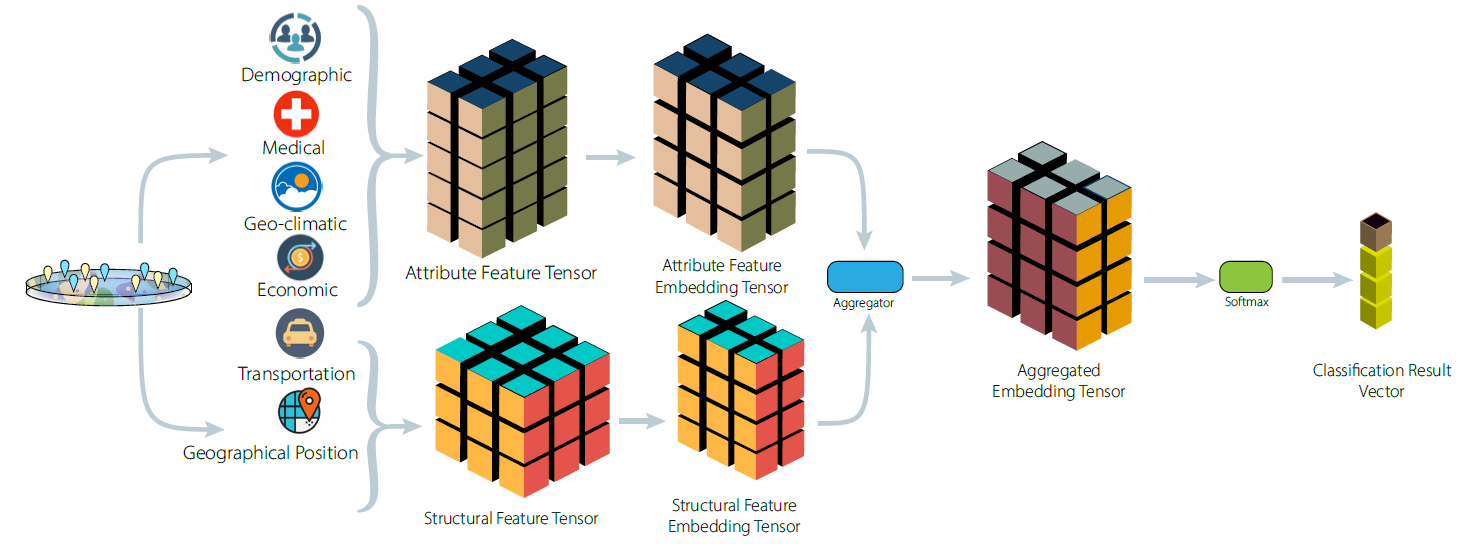}
     \caption{Schematic diagram of PANDORA.}
     \label{fig:flow}
\end{figure*}

\subsection{Node Attribute Features Abstract}
Several factors can cause the COVID-19 outbreak. We consider four critical factors in constructing the node features, i.e., demographic, medical condition, geo-climatic conditions, and economic conditions, respectively.

\subsubsection{Demographic Conditions}
Dalziel et al.~\cite{Dalziel75} proved that some smaller cities (e.g., with less residential density) own longer influenza season. However, in larger and densely populated cities, more-diffuse epidemics generally happen. That is presumably because larger cities own higher rate of personal contact, which makes geo-climatic conditions have a low influence on the transmission of the pandemic. Therefore, we take this fundamental factor as the first considerable feature in predicting infection risk.

\subsubsection{Medical Conditions}
If there are not enough healthcare facilities, the pandemic may also be uncontrollable~\cite{betsch2020social}. Generally speaking, large cities have higher population density, closer transmission distance, higher human mobility, and better medical treatment level and equipment. But small cities have low population density, less contact between people, low human mobility, but insufficient healthcare facilities.

\subsubsection{Geo-climatic Conditions}
It has been studied that different geo-climatic conditions can impact the spread of COVID-19~\cite{zhang2020identifying}. As one of the enveloped viruses, COVID-19 is sensitive to temperature. Geo-climatic conditions will affect the stability of viruses expelled by coughing or sneezing in the external environment.

\subsubsection{Economic Conditions}
Regions with better economic conditions will function well during COVID-19~\cite{bonaccorsi2020economic}. A high level of automatic production, remote office work, and developed urban services guarantee social distancing to a great extent. According to the statistics, economic conditions significantly impact COVID-19 control and prevention. Thus it is significant in predicting the infection risk of a specific region.

We use $demo, med, geo-clim$, and $eco$ to represent the factors mentioned above. The features mentioned above are first processed into discrete numeric data to improve the prediction accuracy and shorten the learning period. The continuous attribute discretization method determines the accuracy of information expression and extraction. However, for Non-Euclidean data such as networked data, the Chi2~\cite{rempala2016double} algorithm outperforms other methods such as Bayesian classifier, Support Vector Machine (SVM), and decision tree in continuous attribute discretization. Through this Chi2 method, we discretize the four mentioned features into 10 categories. On this basis, we transform them into one-hot coding and finally use them as node attribute features. The Chi2 algorithm is based on the $\chi^2$ statistics, with two main procedures:

\begin{enumerate}
\item{First, it defines a high significance level (sigLevel) and sorts all the attribute by their values. Then Chi2 calculates the $\chi^2$ according to Eq.~\eqref{eq:ch2eq}. For each adjacent interval, Chi2 merges the pair of adjacent intervals with the lowest $\chi^2$ value until the $\chi^2$ values of all pairs exceed the parameter determined by sigLevel. The above process uses the decreasing sigLevel until inconsistency rate is exceeded in the discretized data.
\begin{equation}
\chi^2 = \sum_{i=1}^{2}{\sum_{j=1}^k{\frac{(A_{ij}-E_{ij})^2}{E_{ij}}}}
\label{eq:ch2eq}
\end{equation}
wherein, $k$ represents the number of classes, $A_{ij}$ represents the patterns in the $i$th interval, $j$th class, $E_{ij}$ represents the expected frequency of $A_{ij} = R_i\times C_j/N$. Wherein, $R_i$ is the number of examples in $i$th interval, $\sum_{j=1}^{k}{A_{ij}}$, and $C_j$ is the number of examples in $j$th class, $\sum_{i=1}^{m}A_{ij}$}

\item{
Then, the discretization starts with sigLevel determined in the first procedure. After attributes merging process, there is a consistency checking. If the consistency rate does not exceed the target value, it will decrease for the next merging. Otherwise, this attribute will not be involved in further merging until no attribute's value can be merged.
}
\end{enumerate}

\subsection{Network Motif Enumeration}
\begin{figure*}[htbp]
     \centering
     \includegraphics[width=1\textwidth]{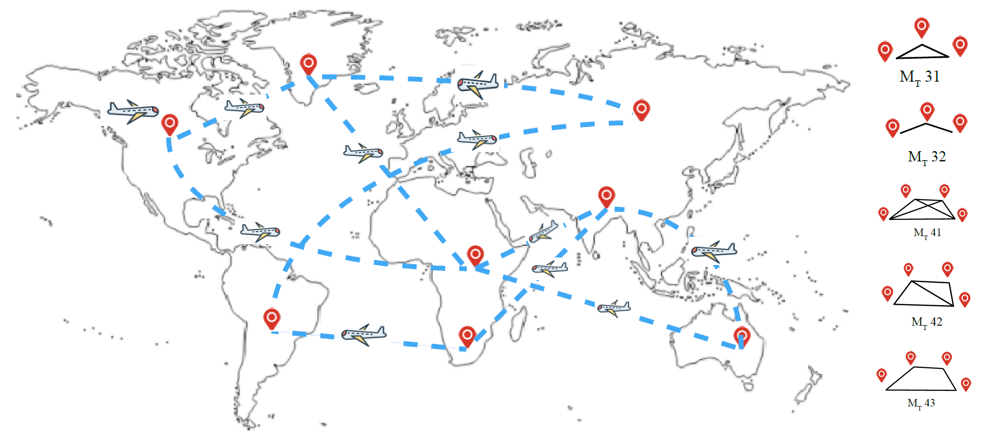}
     \caption{Transmission network motifs employed in this paper.}
     \label{fig:motifs}
\end{figure*}
It is critical to capture transportation relationships among different regions in the transmission of COVID-19. As basic structures of complex networks, network motifs have been widely applied to use as representations of special patterns, and such structure has been proved to be effective in network embedding process~\cite{10.1145/3184558.3186900,rossi2018deep}. Motifs in higher order will lead to high computational complexity, and thus lower-order ones are generally used. In this work, we use nodes of the graph to represent countries or regions, and edges to represent the relationships based on the possible transportation options. And then, we employ 5 kinds of network motifs, which are correspondingly defined as transmission motif $M_T31$, $M_T32$, $M_T41$, $M_T42$, and $M_T43$ (shown in Fig.~\ref{fig:motifs}), to represent the possible transmission relationships between regions. The five employed homogeneous motifs include all kinds of homogeneous three-order motifs and three kinds of four-order motifs. These network substructures have been proved to be effective in formulating the transportation relationships among different regions~\cite{10.1145/3292500.3330869, 9369147}. In addition to the direct connection between the two counties, these 5 motifs may also cause the transmission of the pandemic. In order to characterize the degree of a node involved in particular multivariate relationships, Node Motif Degree ($NMD$) is defined as follows.

\begin{definition}
In the graph $\mathcal{G} = (\mathcal{V},\mathcal{E})$, the node motif degree information $NMD(v)$ of a node $v\in \mathcal{V}$ is defined as the number of five motif $M_T31$, $M_T32$, $M_T4$1,  $M_T42$ and $M_T43$, whose constituent nodes include the node $v$.
\end{definition}

To count the number of five motifs, we use a neighborhood search algorithm to calculate the number~\cite{yu2020motif} quickly. For every node $a$ in the network, set $N_a$ as the set of its neighborhoods. And for every node $b\in N_a$, set $N_b$ as the set of its neighborhoods. Then record the intersection node in $N_a$ and $N_b$ as $Inse$. Then there are some distinct ways to count different kinds of motifs.

\begin{enumerate}
\item{$M_T31$:} Since $M_T31$ is fully-connected, all nodes in $Inse$ are involved in $M_T31$ motif structure.
\item{$M_T32$:} $M_T32$ is a different structure, which is not fully-connected. The node in $N_a$ or $N_b$ but not in $Inse$ accords with the structure in $M_T32$, and this will be recorded by VFlag.
\item{$M_T41$\&$M_T42$:} In the calculation of $M_T41$, we need to add the neighborhoods of every node $c\in Inse$ and define it with $N_c$. For every node in $N_c\in Inse$, the number of $M_T41$ will be added. Conversely, if it is in $N_a$ or $N_b$, i.e., the node is connected to $a$ or $b$, the number of $M_T42$ will be added.
\item{$M_T43$:} When the fourth node is connected to $b$ and $c$ and not connected to $a$, the number of $M_T43$ will be added.
    \end{enumerate}

Through the definition and the above neighborhood search algorithm, we get the motif feature matrix and add it to the construction of the structural attribute.

\begin{table*}[b]
  \centering
  \caption{Detailed information of data sets.}
    \begin{tabular}{ccccc}
    \toprule
   Factor & Data Set & Node/Edge & Scale & Sources \\
    \midrule
    Economic & Unemployment Rate & Node  & 3,234  & Centers for Disease Control and Prevention (CDC) \\
    Mobility & Google Mobility & Node  & 90,552 & Google \\
    Medical & ICU Beds/Deaths Rate & Node  & 3,234  & Kaiser Health News (KHN) \\
    Weather & Temperature & Node  & 45,276 & National Oceanic and Atmospheric Administration (NOAA) \\
    Traffic & Flight & Edge  & 803   & Transportation Security Administration (TSA) \\
    Location & Adjacent & Edge  & 19,352 & The United States Census Bureau (USCB) \\
    \bottomrule
    \end{tabular}
  \label{tab:data}
\end{table*}

\subsection{Graph Embedding}
Since the $\mathbf{AET}$ and $\mathbf{AFT}$ may not have the same scale, blindly connecting them may cause the data loss. Graph embedding can solve the problem that two tensors are not on the same scale. In our model, we choose Graph Convolutional Network to implement the graph embedding process. Based on graph data specification, graphs are a meaningful and understandable representation of data. So that, the adjacency matrix is defined as shown in Eq.~\eqref{eq:adj}.
\begin{equation}
\label{eq:adj}
A_{ij} =
\begin{cases}
1& \text{$if \{v_i, v_j\}\in \mathcal{E}$ and $i \neq j $}\\
0& \text{otherwise}
\end{cases}
\end{equation}
The graph representation using the adjacency matrix has the problem of computational efficiency. Adjacency matrix is a  $|\mathcal{V}|*|\mathcal{V}|$ matrix to represent a graph. $|\mathcal{V}|$ depicts the nodes in the graph. As the number of nodes increases, the space required for this method increase exponentially. That happens when we employ the adjacency matrix to represent a large graph. At the same time, the vast majority of the adjacency matrix is 0, so the fast and effective learning methods are difficult to be applied due to the sparsity of the data. Graph embedding refers to learning the low dimensional vector of nodes in the network, and it needs to capture the graph topology, vertex-to-vertex relationship, and other information. The core idea of the graph embedding process is to learn a function $f$ by aggregating the attribute of nodes $v_i$  and its neighbors. A robust mathematical representation of the graph is the normalized graph Laplacian matrix, which is shown in Eq.~\eqref{eq:laplacian}.

\begin{equation}
\label{eq:laplacian}
\mathbf{L} = \mathbf{I_n} - \mathbf{D}^{-\frac{1}{2}} \mathbf{A} \mathbf{D}^{-\frac{1}{2}}
\end{equation}
wherein, $\mathbf{A}$ is the adjacency matrix of the graph, $\mathbf{I_n}$ is the identity matrix, and $D$ is the diagonal matrix of node degree, $\mathbf{D_{ii}} = \sum{(\mathbf{A_{i,j}})}$. The normalized Laplacian matrix has the property of symmetric positive semidefinite, therefore, it can be decomposed into Eq.~\eqref{eq:laplacian2}:
\begin {equation}
\label{eq:laplacian2}
\mathbf{L} = \mathbf{U} \Lambda \mathbf{U^T}
\end{equation}
Among them, $\mathbf{U} = [u_0,u_1,\cdots,u_{n-1}]\in \mathbb{R^{N*N}}$ is the matrix composed of the eigenvectors of $L$, $\mathbf{\Lambda}$ is the diagonal matrix of eigenvalues, $\mathbf{\Lambda_{ii}} = \lambda_i$. In graph signal processing, a graph signal $x\in R^N$ is a feature vector composed of graph nodes, where $x_i$ represents the value of node $i$. For signal $x$, the Fourier transform on the graph can be defined as shown in Eq.~\eqref{eq:fourier}.
\begin{equation}
\label{eq:fourier}
\mathcal{F}(x) = \mathbf{U^T}x
\end{equation}
Inverse Fourier transform is defined as shown in Eq.~\eqref{eq:inversefourier}:
\begin{equation}
\label{eq:inversefourier}
\mathcal{F}^{-1}(\hat{a}) = \mathbf{U}\hat{x}
\end{equation}
where, $\hat{x}$ is the result of Fourier transform, The element of the transformed signal $\hat{x}$ is the coordinate of the new spatial signal, so the input signal can be expressed as shown in Eq.~\eqref{eq:inputsignal}:
\begin{equation}
\label{eq:inputsignal}
x = \sum_i{\hat{x}_iu_i}
\end{equation}
For the input signal $x$, the graph convolution can be defined as shown in Eq.~\eqref{eq:gcnconvolution}.
\begin{equation}
\label{eq:gcnconvolution}
x\times_Gg = \mathcal{F}^{-1}(\mathcal{F}(x)\odot\mathcal{F}(g))\\
=\mathbf{U}(\mathbf{U^T} x \odot \mathbf{U^T}g)
\end{equation}
Where, g$\in \mathbb{R}^N$ is the filter and $\odot$ represents the aggregation, and the $\times_{g}$ represents the graph convolution operation. The embedding process can be defined as shown in Eq.~\eqref{eq:GCNembedding}.
\begin{equation}
\label{eq:GCNembedding}
g_\theta\times_{G}x = \theta(\mathbf{\bar D^{-\frac{1}{2}}}\mathbf{\bar A}\mathbf{ \bar D^{-\frac{1}{2}}})x
\end{equation}

\subsection{The Overall Framework}
To better classify infection risk levels, we consider the attribute information such as medical conditions, economic conditions, and the number of infected people. The Attribute Feature Tensor (AFT) is generated based on these factors. We then calculate 5 kinds of transmission motifs degree information of each node in the network. All of the information will be integrated with the initial county adjacency information and transportation information. Therefore, the Structural Feature Tensor (SFT) is then obtained. Our method uses Graph Convolutional Networks (GCN) to get the representations of AFT and SFT. Since the two tensors may differ in scale, blindly connecting the two tensors may cause data loss. Therefore, the network embedding process is employed for both AFT and SFT to achieve tensors in the same dimension.

\begin{algorithm}[htb]
  \caption{Framework of PANDORA}
  \label{alg:PANDORA}
    \KwIn {A graph $\mathcal{G} = (\mathcal{E},\mathcal{V})$.}
    \KwOut {Result Vector.}
    \For{each $\forall v\in \mathcal{V}$ }{
    Extracts the attribute information and calculates five kinds of node motif degree information.\\
    Integrate it into $\mathbf{AFT}$ and $\mathbf{SFT}$, respectively.\\
    }
    \For{each $\forall v\in \mathcal{V}$}{
    $\mathbf{H^{0}_s}$ is $\mathbf{SFT}$ and $\mathbf{H^{0}_a}$ is$\mathbf{AFT}$.\\
     $\mathbf{H^{i+1}_s}  =  \theta(\mathbf{\bar D^{-\frac{1}{2}}} \mathbf{\bar A} \mathbf{\bar D^{-\frac{1}{2}}})\mathbf{H^{i}_s}$\\
     Integrate it into $\mathbf{AEFT}$ and $\mathbf{SFET}$, respectively.\\
     $\mathbf{H^{i+1}_a}  =  \theta(\mathbf{\bar D^{-\frac{1}{2}}} \mathbf{\bar A} \mathbf{\bar D^{-\frac{1}{2}}})\mathbf{H^{i}_a}$\\
     Integrate it into $\mathbf{AEFT}$ and $\mathbf{SFET}$, respectively.\\
    }
    \For{each $\forall v\in \mathcal{V}$}{
      \If { Using Hadamard aggregation method.}{
       $\mathbf{AET} = (\mathbf{AEFT}\odot \mathbf{SFET})$\\
     }
      \If { Using Summation aggregation method.}{
    $\mathbf{AET} = \mathbf{AEFT}+\mathbf{SFET}$\\
     }
     \If { Using Connection aggregation method.}{
     $\mathbf{AET} = \left[\mathbf{AEFT},\mathbf{SFET}\right]$\\
     }
     }
     Get $\mathbf{AET}$.\\
     Result Vector $\gets Softmax(\mathbf{AET})$  \\
    \Return Result Vector;
  \end{algorithm}



The general process of the proposed model is shown in Fig.~\ref{fig:flow}. We can then obtain the Attribute Feature Embedding Tensor ($\mathbf{AFET}$) and Structural Feature Embedding Tensor ($\mathbf{SFET}$). In order to aggregate two embedding tensors, three aggregation methods are proposed, i.e., Hadamard Aggregation (HA), Summation Aggregation (SA), and Connection Aggregation (CA). HA is the dot multiplication of two embedding tensors; SA is the added value of two embedding tensors, and CA is the connection of two embedding tensors. In terms of node information, HA and SA method may lose a part of the information of the features but CA retain all the information in a connected way. In terms of time consumption, HA and SA method will not change the size of the input tensor, but CA will double the size of the input tensor.
We use classification tasks to evaluate the network representation learning results. In the classification tasks, we add the Softmax function to generate the node classification result vector, which is shown in Eq.~\eqref{eq:sm}.
\begin{equation}
\label{eq:sm}
S_j = \frac{e^{a_j}}{\sum_{k = 1}^K{e^{a_k}}}
\end{equation}
wherein, $K$ represents the number of the classification and $a_j$, and $a_k$ represents the $j$-th or $k$-th value of in the vector. The above is the process of PANDORA, which can be expressed as shown in Eq.~\eqref{eq:PANDORA}. In the dynamic graph, we use the PANDORA embedding method at each timestamp and finally add the results. The process is shown in Eq.~\eqref{eq:PANDORADY}.

\begin{equation}
\hat Y = SM(Agg(GCN(\mathbf{AFT}),GCN(\mathbf{SFT}))\cdot\Theta)
\label{eq:PANDORA}
\end{equation}
\begin{equation}
\hat Y = SM(\sum_t^T Agg(GCN(\mathbf{AFT_t}),GCN(\mathbf{SFT_t}))\cdot\Theta)
\label{eq:PANDORADY}
\end{equation}
wherein, $ATF_t$ and $SFT_t$ are the structural feature tensor and attribute feature tensor on timestamp, $GCN(\cdot)$ represents the GCN iteration process, and $Agg(\cdot)$ represents the aggregation process, $SM(\cdot)$ represents the Softmax function.
We employ the cross-entropy error as the loss function of PANDORA, and the specific error function of one data is shown in Eq.~\eqref{eq:loss}.
\begin{equation}
L_i = -\sum_{l=1}^{\#label}Y_{il}\cdot\ln(\hat{Y_{il})}
\label{eq:loss}
\end{equation}
wherein, $Y$ represents the true node classification label, and $\hat{Y}$ represents the label predicted by the model.

\begin{figure*}[tp]
  \centering
  \subfigure[Confirmed cases]
{
    \centering
    \includegraphics[width=.47\textwidth]{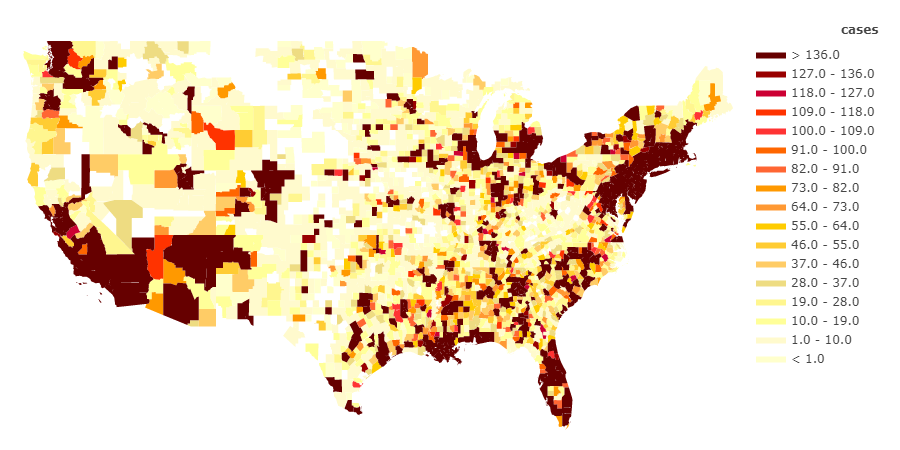}
    \label{fig:cases}
}
  \subfigure[Medical level]{
    \centering
    \includegraphics[width=.47\textwidth]{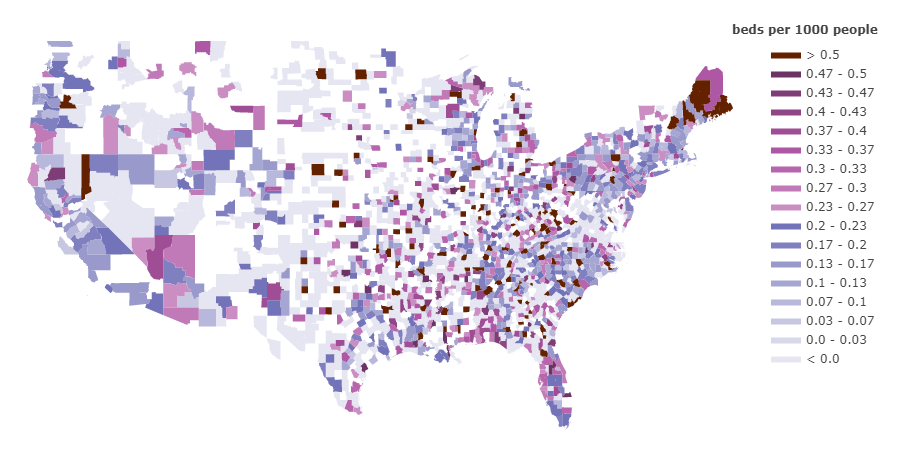}
    \label{fig:medical}
}
 \subfigure[Economic condition]{
    \centering
    \includegraphics[width=.47\textwidth]{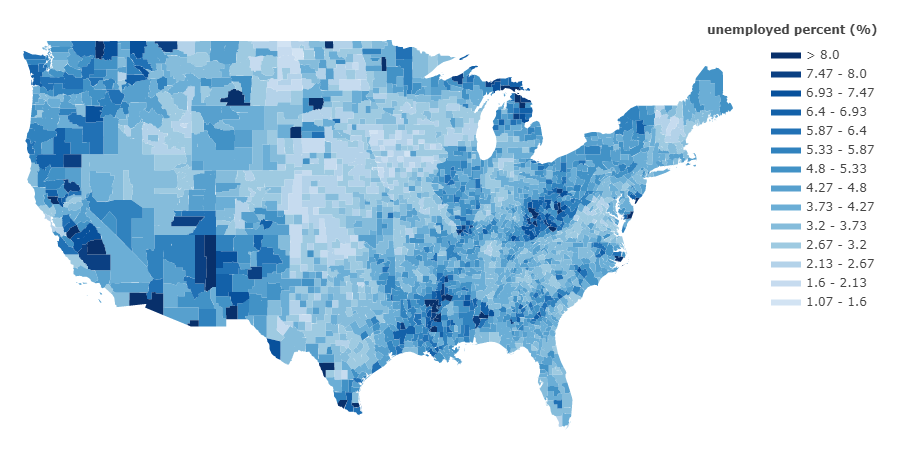}
    \label{fig:eco}
}
  \subfigure[Mobility]{
    \centering
    \includegraphics[width=.47\textwidth]{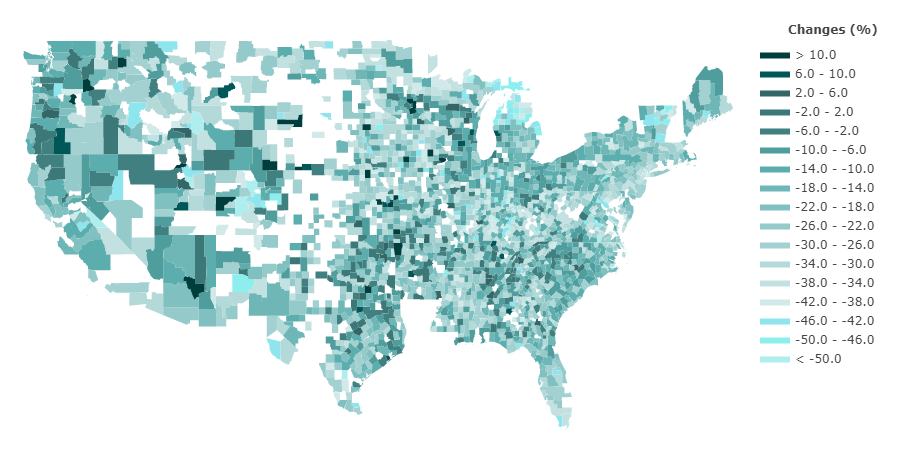}
    \label{fig:mobility}
}
    \caption{Distributions of four kinds of data.}
    \label{fig:datasetsvis}
\end{figure*}
\section{Experiments}
\label{sec:exp}

\subsection{Data Sets}

Rich data can make a huge contribution to the fight against COVID-19. The number of cases and mortality in time series are the data that most researchers focus on. On this basis, we have additionally selected the following data sets shown in Table~\ref{tab:data} to help with the generation of our network to better implement the experiments.
\begin{figure*}[!htb]
  \centering
  \subfigure[Skewed distribution of medical level feature data]{
\centering
    \includegraphics[width=.22\textwidth]{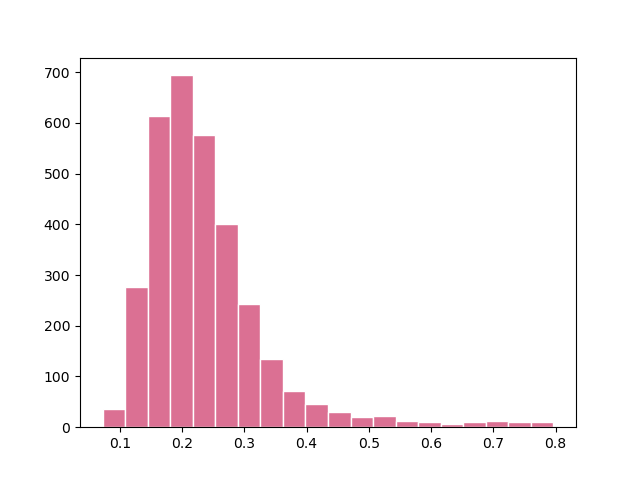}
    \label{fig:skdismedical}
}
 \subfigure[Normal distribution of medical level feature data]{
\centering
    \includegraphics[width=.22\textwidth]{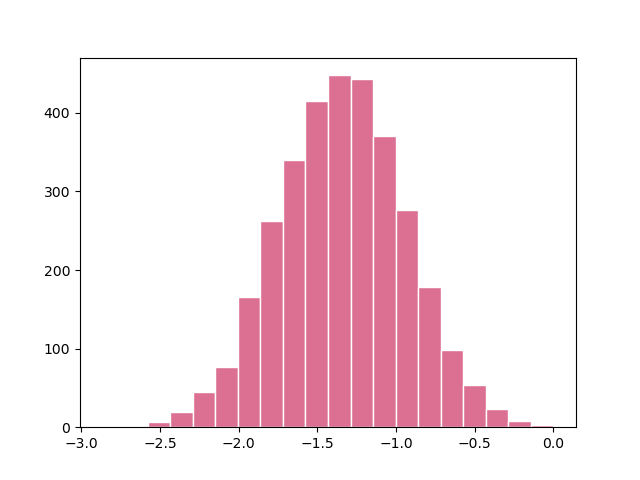}
    \label{fig:normaldismedical}
}
  \subfigure[Skewed distribution of economic condition feature data]{
\centering
    \includegraphics[width=.22\textwidth]{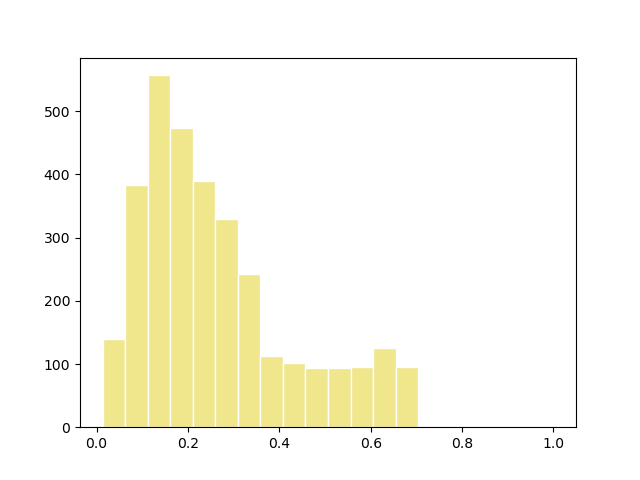}
    \label{fig:skdiseconomic}
}
 \subfigure[Normal distribution of economic condition feature data]{
\centering
    \includegraphics[width=.22\textwidth]{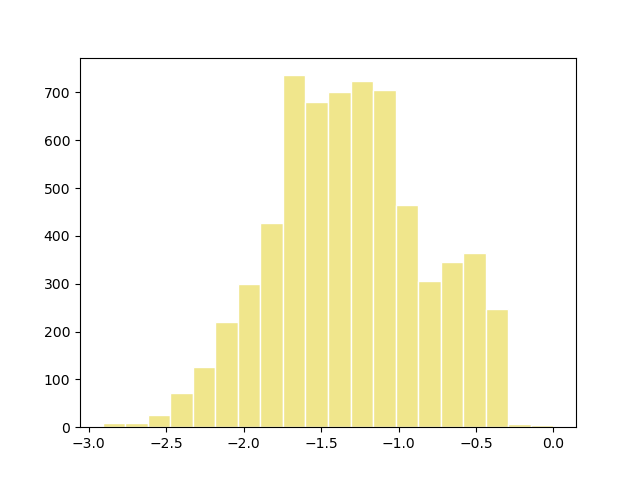}
    \label{fig:normaldiseconomic}
}
    \caption{Distribution normalization in pre-processing data.}
    \label{fig:normal}
\end{figure*}

The data are composed of two parts, including edge information and node information. We specifically select information of 3,234 (of a total 3,243) counties in the United States as the network's nodes. 9 counties do not have the specific information mentioned in Table~\ref{tab:data}, so those counties are not selected. In terms of structural attributes, we obtain the flight information from March 1st, 2020, to April 15th, 2020. This data set provides the list of positive inspectors, airport staff and where they work, and the date they finished their work. We then process the above data and integrate related data into distinct counties. In addition, we also considered that the geographical location of adjacent counties might cause the transmission of the pandemic, so we obtained the information of neighboring counties from Census\footnote{https://www.census.gov/geographies/reference-files/2010/geo/county-adjacency.html/}. We then integrated flight information and adjacent information as the edges in the network. In terms of node attributes, we obtained the numbers of confirmed cases, deaths, and the population from the New York Times\footnote{https://github.com/nytimes/covid-19-data/}. The number of confirmed cases in counties is shown in Fig.~\ref{fig:cases}. Besides, we also select the number of ICU beds in each county from KHN\footnote{https://khn.org/news/as-coronavirus-spreads-widely-millions-of-older-americans-live-in-counties-with-no-icu-beds/\#lookup/} to represent the medical level of each county in coping with the pandemic. Considering that using only the number of ICU beds to represent the medical level of a county is inadequate, we also use the ratio of deaths to confirmed cases per county as the second indicator of the medical level from the New York Times\footnote{https://github.com/nytimes/covid-19-data/}. In order to ensure the uniformity of data in each county, we processed the data into the number of beds per 1,000 people. Details are shown in Fig.~\ref{fig:medical}.
In the representation of geo-climatic, we collect climate data from NOAA\footnote{https://www.ncei.noaa.gov/data/global-summary-of-the-day/access/2020/}, and the temperature is used in the experiment. We also obtain the unemployment rate from Disease Control and Prevention\footnote{https://www.cdc.gov/dhdsp/maps/sd\_unemployment.htm/} as a representative measurement of the economic situation for each county (shown in Fig.~\ref{fig:eco}). Moreover, the mobility data are also collected from Google\footnote{https://www.google.com/covid19/mobility/} as the reference for the spread of COVID-19, and then these data are further processed by calculating the mean values for each county. Details are shown in Fig.~\ref{fig:mobility}.
We choose 60\% as a training set, 20\% as a validation set, and 20\% as a test set. All nodes in the data set are labeled.

\subsection{Data Pre-processing}




We preprocess the network to clearly express the characteristics of each node and obtain better experimental results. Firstly, we generate 3,234 counties in the United States and use the information of the corresponding state.

There are nearly 700 counties in the scattered distribution of the western United States, which occupies more than 20\% of all counties. Counties in the western U.S. are generally in scattered distribution due to the underdevelopment of economics, leading to the whole data appearing in a skewed distribution. Therefore, we use the box-cox method to convert all data distribution to normal distribution (shown in Fig.~\ref{fig:normal}). On this basis, we normalize all the data to speed up the training process.

Counties with adequate flight information and geographically bounded are regarded as connected in the network. The edge in the network represents there are flights between two certain counties, or the two counties are adjacent.

Based on the definition by Centers for Disease Control and Prevention (CDC)~\footnote{https://www.cdc.gov/coronavirus/2019-ncov/travelers/how-level-is-determined.html/}, the increased number of confirmed cases is employed to classify infection regions into 4 different risk levels, i.e., risk-free, low-risk, medium-risk, and high-risk.


According to the standard mentioned above, we label the counties in our data set to verify the classification results of PANDORA. In general, there are 700 risk-free counties, 1,183 low-risk counties, 1,048 medium-risk counties, and 210 high-risk counties.

\begin{figure}[!htbp]
     \centering
     \includegraphics[width=.4\textwidth]{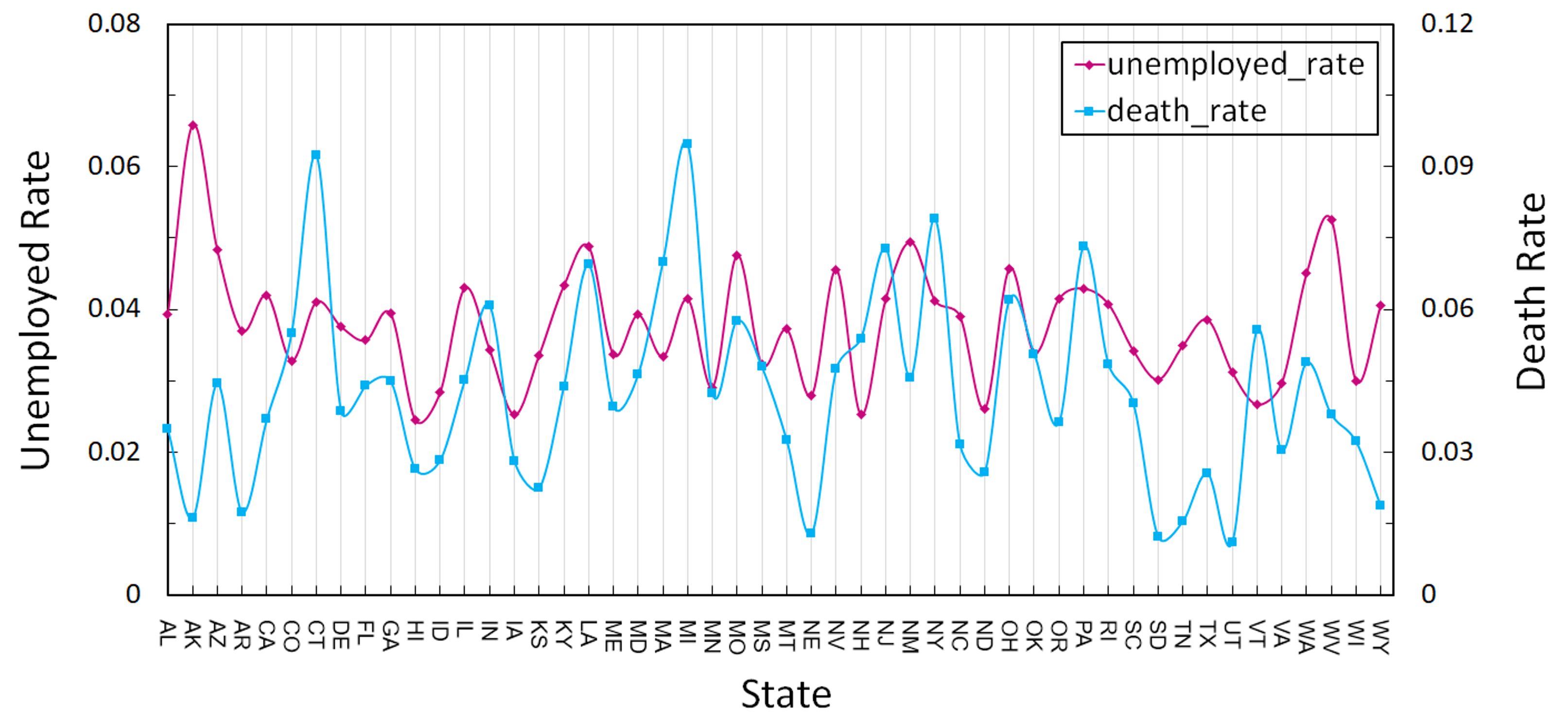}
     \caption{Analysis of death rate of COVID-19 and unemployment rate from January 21st, 2020 to June 2nd, 2020.} \label{death-economic}
\end{figure}

\begin{figure}[!htbp]
	\centering
	\includegraphics[width=.4\textwidth]{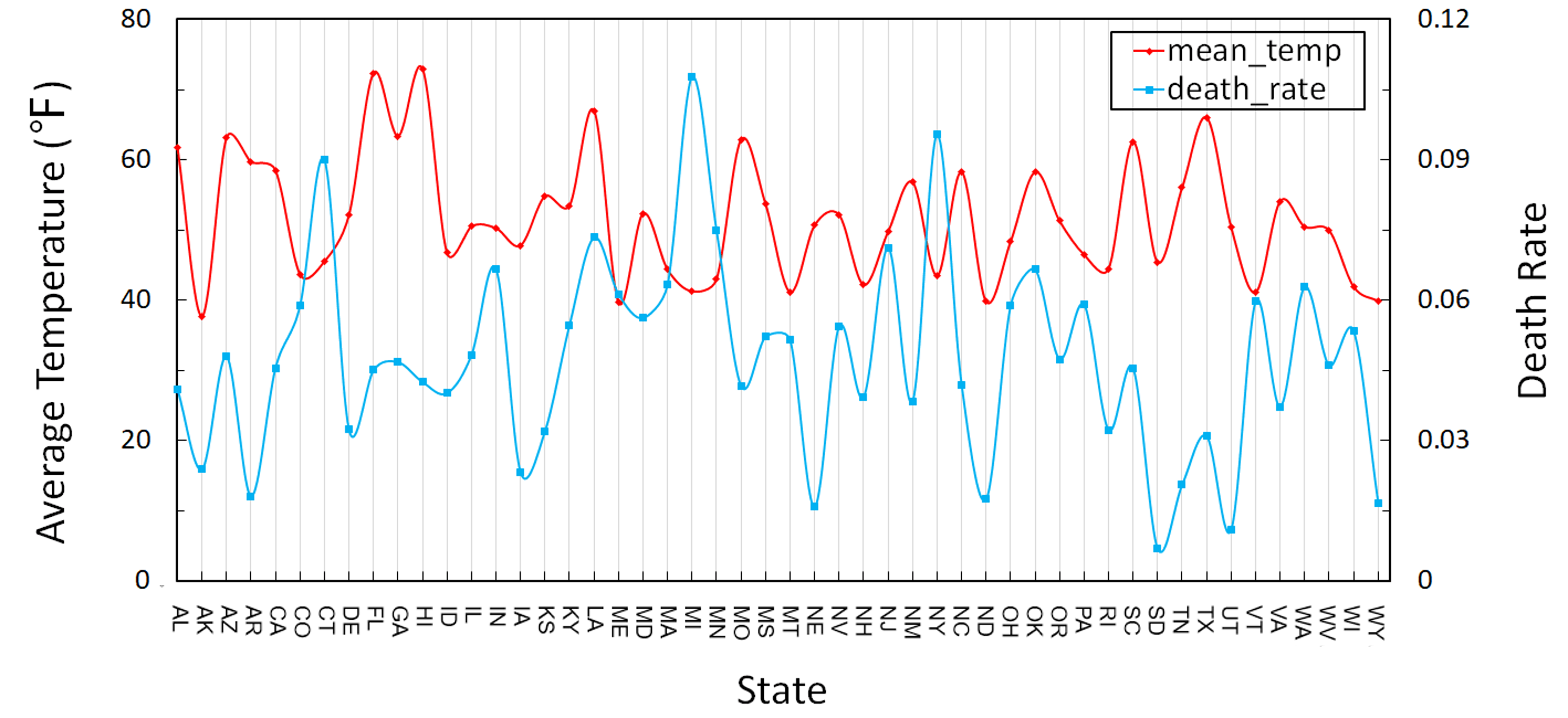}
	\caption{Analysis of death rate of COVID-19 and monthly average temperature from April 1st, 2020 to April 30th, 2020.}
     \label{death-temperature}
\end{figure}

\begin{figure}[!htbp]
	\centering
	\includegraphics[width=.4\textwidth]{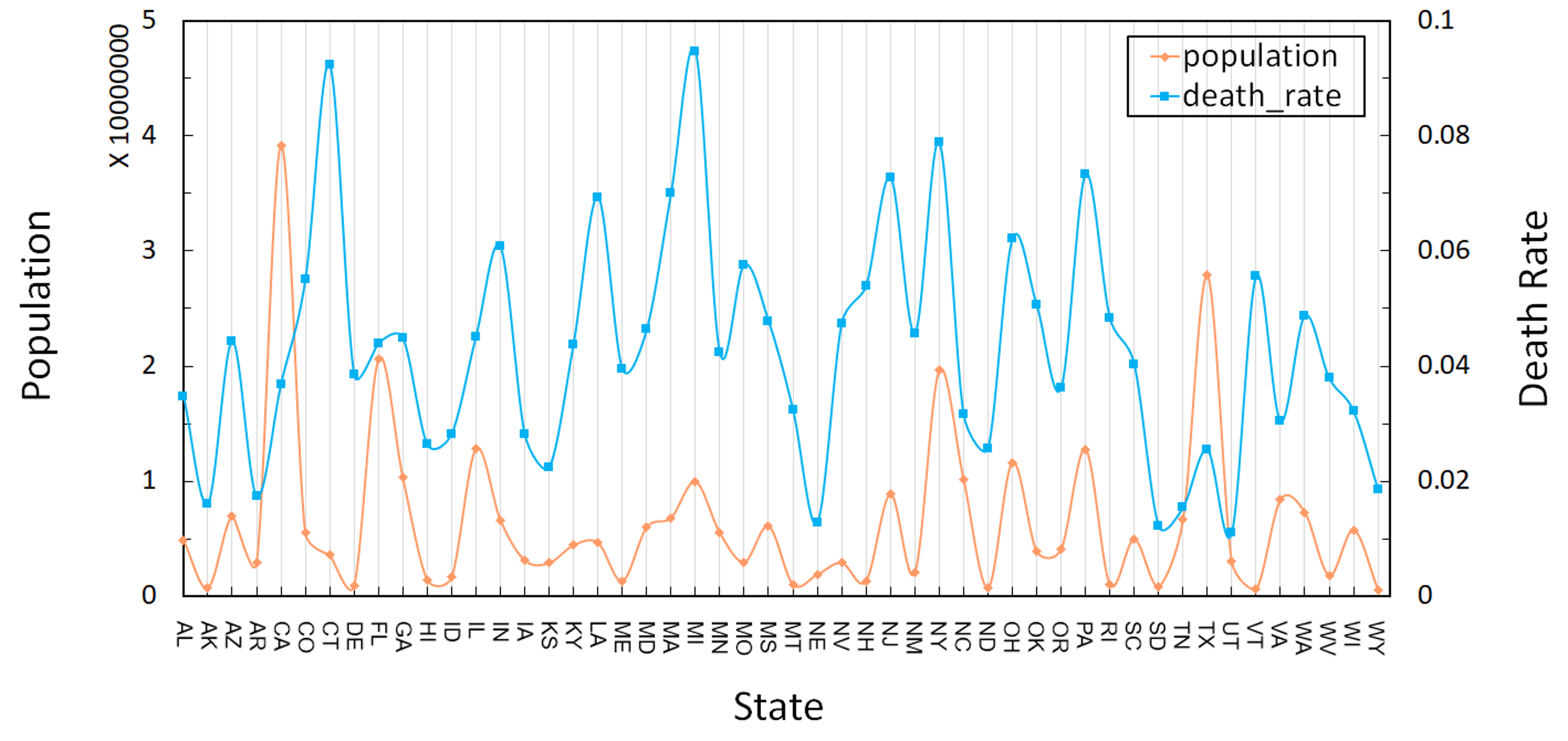}
	\caption{Analysis of death rate of COVID-19 and population size from April 1st, 2020 to April 30th, 2020.}
     \label{death-population}
\end{figure}
\begin{table*}[hb]
  \centering
  \caption{Comparison in terms of accuracy.}
    \begin{tabular}{ccccc}
	    \toprule
	    Hyperparameters & Methods & Accuracy& Iteration &Highest Accuracy\\
	    \midrule
	    \multirow{4}[2]{*}{$\alpha = 0.01, {max\_epoch = 300}$} & GCN   & 80.50\%&300&80.83\% \\
 & PANDORA-HA & 82.00\%&179& 82.50\%\\
& PANDORA-SU & \textbf{82.17\%} &127&82.17\%\\
 & PANDORA-CO & 81.67\% &115&81.67\%\\
	    \midrule
	    \multirow{4}[2]{*}{$\alpha = 0.003, {max\_epoch = 500}$} & GCN   & 79.67\%&500&79.67\% \\
	          & PANDORA-HA & 81.00\%&388 &81.33\%\\
	          & PANDORA-SU & 81.67\% &319&81.83\%\\
	          & PANDORA-CO & \textbf{81.17\%}&243 &81.50\%\\
	    \midrule
	    \multirow{4}[2]{*}{$\alpha = 0.0008, {max\_epoch = 1000}$} & GCN   & 77.83\%&1000&78.00\% \\
	          & PANDORA-HA & 79.17\% &1000&79.33\%\\
	          & PANDORA-SU & \textbf{81.50\%} &962&81.83\%\\
	          & PANDORA-CO & 81.00\%&768&81.00\%\\
	    \bottomrule
	    \end{tabular}
  \label{tab:acc}
\end{table*}
We also build a dynamic graph in addition to constructing a static graph that aggregates all factors into a vector as node features. Herein, we consider three factors as the node feature in each timestamp: confirmed cases, mobility, and climatic conditions. Since medical and economic levels have not changed in terms of what we used, we did not consider these two factors in the dynamic graph.

\subsection{Correlation Analysis}
\label{sec:cor}

This section visualizes the collected data and analyzes the correlation between COVID-19 death rate and temperature, economy, and population size. Herein, we specifically use the death rate instead of the infection rate. It is because preventing COVID-19 infection has become a global consensus. Therefore, isolating policies are generally implemented. Compared to the infection rate, the death rate can reflect whether the medical system of a certain region is broken down or not. To better represent these data, we generally use 50 states of the United States to show the general correlation.

To analyze the relationship between death rate and economy, we select the unemployment rate as a representative measure of each state's economy. Fig.~\ref{death-economic} shows the curves of the death rate and unemployment rate of COVID-19 from January 21st, 2020, to June 2nd, 2020. The horizontal axis represents the 50 states, and the vertical axis implies the rate (mortality rate and unemployment rate). It can be seen from the figure that the overall trends of the two variables are similar. Therefore, the unemployment rate is used as a reference for the state's economic level.

Fig.~\ref{death-temperature} shows the curve of the death rate of COVID-19 and the monthly average temperature in the United States from April 1st, 2020, to April 30th, 2020. The horizontal axis represents the 50 states of the United States, and the vertical axis implies the monthly average temperature of each state. The secondary vertical axis signifies the death rate of each state. Some existing studies have also demonstrated the effect of temperature on the spread of COVID-19 through accurate analysis.

Fig. ~\ref{death-population} shows the curve of the death rate and population size of COVID-19 from January 21st, 2020, to June 2nd, 2020. The horizontal axis represents the 50 states in the United States, and the vertical axis implies the total population of each state. The secondary vertical axis signifies the death rate of each state. The figure shows that the trend of the two is basically the same (except Colorado (CO) and Connecticut (CT)), so the death rate and the population size have a strong correlation. Therefore, crowd gathering should be avoided in populous regions.

\subsection{Baselines}
We use GCN to evaluate the accuracy and efficiency of our method on the static graph. On the dynamic graph, we use the following dynamic graph representation learning as baselines.
\begin{itemize}
\item{\textbf{ST-GCN}~\cite{10.5555/3304222.3304273} tackles the time series prediction problem in traffic domain. Complete convolutional structure is established in ST-GCN.}
\item{\textbf{Graph WaveNet}~\cite{10.5555/3367243.3367303} tackles two problems, i.e., the supplementation of graph structure information and long-term dependence, from both spatio and temporal perspectives.}
\item{\textbf{AGCRN}~\cite{bai2020adaptive} does not apply the pre-defined graph structure, but learns the specific mode of each node.}
\item{\textbf{CURB-GAN}~\cite{10.1145/3394486.3403127} provides estimation in a continuous period of time based on different requirements.}
\end{itemize}

\begin{figure*}[htbp]
  \centering
  \label{fig:accandloss}
 \subfigure[Change in Accuracy on TRS]{
     \label{fig:accTRS}
    \includegraphics[width=.3\textwidth]{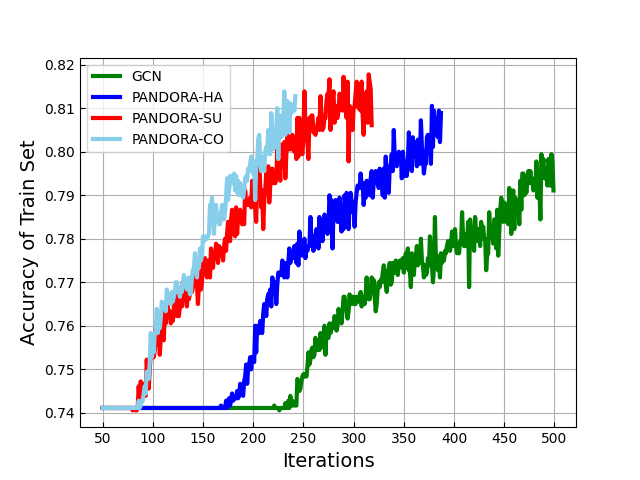}
}
\subfigure[Change in Accuracy on VAS]{
\label{fig:accVAS}
    \includegraphics[width=.3\textwidth]{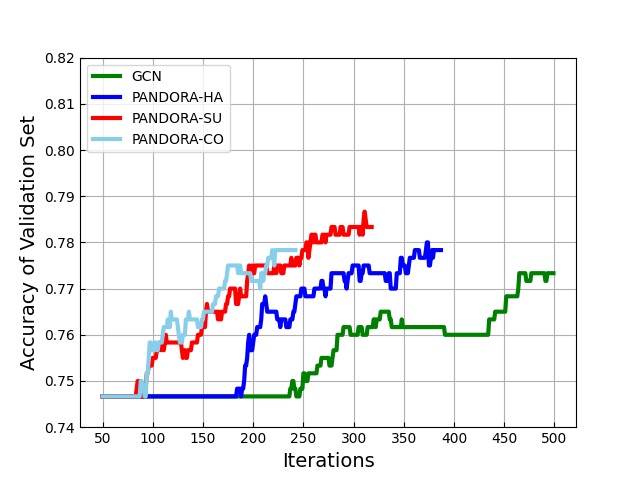}
    
}
  \subfigure[Change in Accuracy on TES]{
     \label{fig:accTES}
    \includegraphics[width=.3\textwidth]{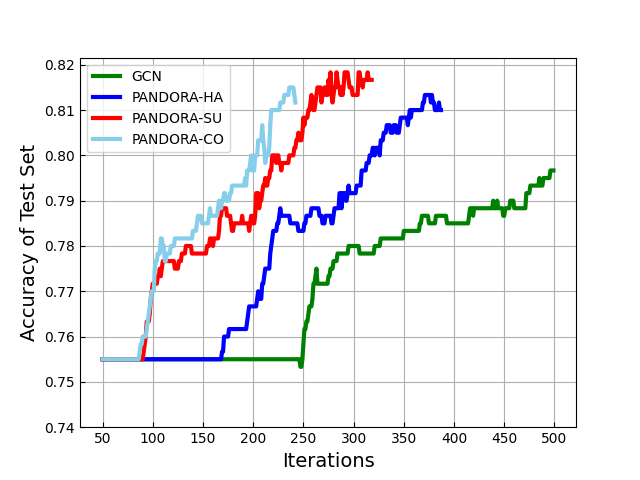}
    
}\\
\subfigure[Change in LOSS on TRS]{
     \label{fig:lossTRS}
    \includegraphics[width=.3\textwidth]{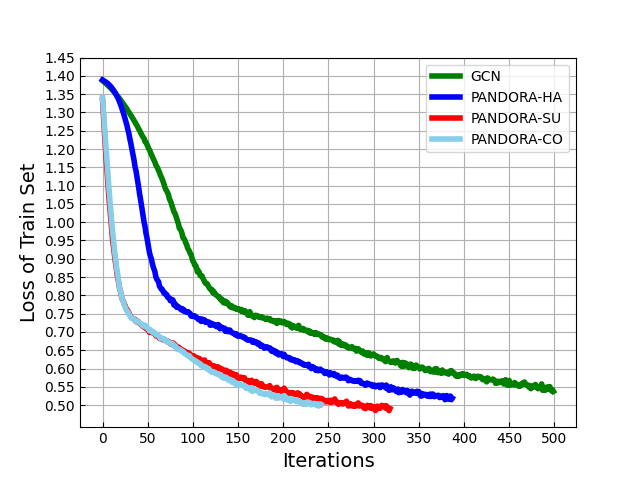}
    
}
  \subfigure[Change in LOSS on VAS]{
  \label{fig:lossVAS}
    \includegraphics[width=.3\textwidth]{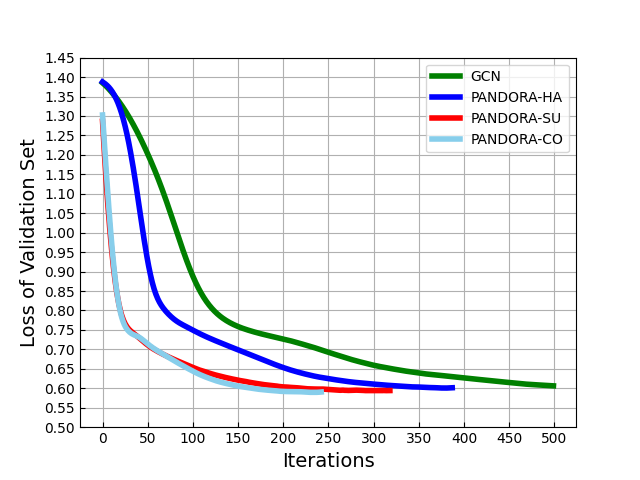}
   
}
\subfigure[Change in LOSS on TES]{
\label{fig:lossTES}
    \includegraphics[width=.3\textwidth]{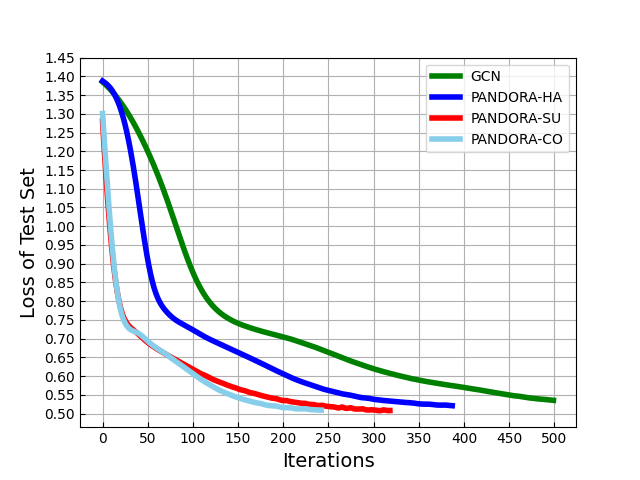}
    
}
  \caption{Comparison of PANDORA-HA, PANDORA-SU, PANDORA-CO, and GCN in convergence speed. The hyperparameters are set to be $\alpha$ = 0.003, max\_epoch = 500, and ED = 64, respectively.}    
\end{figure*}

\begin{figure*}[!ht]
	\centering
	\includegraphics[width=1\textwidth]{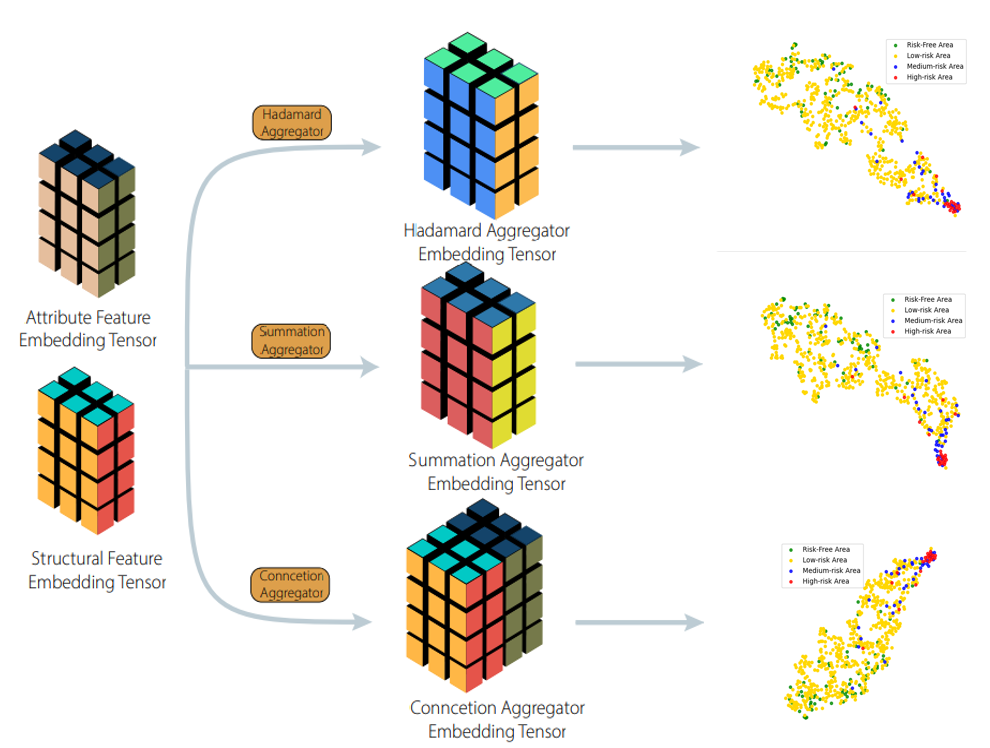}
	\caption{Visualization of results of three models. The hyperparameters of the running environment are set to $\alpha$ = 0.01, max\_epoch = 300, and ED = 64.}
	\label{fig:TSNE}
\end{figure*}

\subsection{Experimental Results}
\subsubsection{Static Graphs}
We set 3 groups of hyperparameters that differ in the learning rate $\alpha$ and the maximum number of iterations $max\_epoch$. The 3 groups of hyperparameters are [0.01, 300], [0.003, 500], [0.0008, 1000]. PANDORA with 3 different aggregators (i.e., PANDORA-HA, PANDORA-SU, and PANDORA-CO) and GCN are then iterated using these 3 groups of hyperparameters. The experimental results of accuracy are shown in Table~\ref{tab:acc}. It can be seen that the proposed PANDORA model outperforms GCN on all three sets of hyperparameters. Specifically, PANDORA-HA performs the best with the accuracy of 82.17\% when $\alpha=0.01$ and $max\_epoch=300$. PANDORA-SU performs the best with the accuracy of 81.67\% when $\alpha=0.003$ and $max\_epoch=500$. It also performs the best in the last group, and all proposed methods perform better than GCN.

We then implement comparative experiments to verify the efficiency of the proposed method and GCN. We iterate the four models to convergence with a learning rate $\alpha$ of 0.003. To ensure all methods fall in convergence, we set the max epoch as 2,000. The embedding dimension (ED) is 64 (128 in the connection aggregation method). To ensure all methods will fall in convergence, we set the iteration time as 2,000. We use two-layer GCNs and choose ReLU as the activation function in the first layer. The input dimension of node features is 600 on the static graph and 4 in each timestamp on the dynamic graph. And when adapting the summation and Hadamard aggregation methods, the input dimension of the second layer is set to be 64 and 128 in the connection method. The output dimension is 5.

The experimental results are shown in Table~\ref{tab:efficiency}. We use Accuracy, Iteration time to convergence, Average Single Training Time \textbf{(ASTT)}, Overall Iteration Time \textbf{(OIT)}, and Test Running Time \textbf{(TET)}, to measure the convergence speed and training efficiency of the model.

It can be seen from Table~\ref{tab:efficiency}, the proposed PANDORA models perform better than GCN with much less OIT and higher accuracy at the same time. Meanwhile, iteration times that PANDORA models carried out are much less than GCN. GCN uses 645 iteration times. For the proposed PANDORA, PANDORA-CO only uses 243 iteration times. Though PANDORA-HA uses the most iteration times, i.e., 388, it is still 257 times less than GCN. Therefore, PANDORA has an overall better performance compared to GCN.

\begin{table}[!bp]
  \centering
  \caption{Efficiency comparison of GCN, PANDORA-HA, PANDORA-SU, and PANDORA-CO. The hyperparameters are set to $\alpha$ =0.003, max\_epoch = 2000, and ED = 64, respectively.}
\resizebox{0.50\textwidth}{!}{
    \begin{tabular}{cccccc}
    \toprule
    Model & \multicolumn{1}{l}{Accuracy} & \multicolumn{1}{l}{Iteration} & \multicolumn{1}{l}{ASTT(s)} & \multicolumn{1}{l}{OIT(s)} & \multicolumn{1}{l}{TET(s)} \\
    \midrule
    GCN   & 80.67\% & 645   & 0.1273 & 113.7055 & 0.0644 \\
    PANDORA-HA & 81.00\% & 388   & 0.0987 & 50.4906 & 0.0500 \\
    PANDORA-SU & \textbf{81.67}\% & 319   & 0.1012 & 42.5297 & 0.0820 \\
    PANDORA-CO & 81.17\% & 243  & 0.1121 & 35.8710 & 0.0530 \\
    \bottomrule
    \end{tabular}}
  \label{tab:efficiency}
\end{table}

\begin{table}[!htbp]
	\centering
	\caption{Performance Evaluation on Dynamic Graphs}
	\resizebox{0.50\textwidth}{!}{
	\begin{tabular}{ccccc}
		\toprule
		& Accuracy & Macro-Pre & Macro-Rec & Macro-F1 \\
		\midrule
		ST-GCN & 56.46\% & 53.97\% & 59.59\% & 56.78\% \\
		Graph WaveNet & 61.75\% & 59.15\% & 61.49\% & 60.32\% \\
		AGCRN & 63.33\% & 60.57\% & 68.33\% & 64.45\% \\
		CURB-GAN & 68.71\% & 67.33\% & 71.11\% & 69.22\% \\
		\midrule
		PANDORA-HA & \textbf{77.50\%} & \textbf{76.87\%} & 77.73\% & \textbf{77.30\%} \\
		PANDORA-CO & 77.00\% & 76.56\% & \textbf{77.86\%} & 77.21\% \\
		PANDORA-SU & 77.33\% & 76.72\% & 77.76\% & 77.24\% \\
		\bottomrule
	\end{tabular}}%
	\label{tab:res_in_dy}%
\end{table}%

We then compare the convergence speed of PANDORA and GCN. The comparison results are shown in Fig.~\ref{fig:accandloss}. Therein, $\alpha=0.003$. The iteration process will continue until one of the four models converges. All of the four methods are respectively implemented in three different sets, including training set (TRS), validation set (VAS), and test set (TES). Accuracy and loss of the four models are calculated respectively on the sets mentioned above. In Fig.~\ref{fig:accandloss}, the horizontal axes are iteration times in all of the six subgraphs. Fig. 9~\subref{fig:accTRS} to Fig. 9~\subref{fig:accTES} show the comparison results of accuracy and Fig. 9~\subref{fig:lossTRS} to Fig. 9~\subref{fig:lossTES} show the change in loss in each set. It can be seen clearly from these figures that the convergence speed of PANDORA is significantly higher than that of GCN. The result shows that PANDORA-CO performs the best compared to PANDORA-HA and PANDORA-SU. That is because all node information is preserved in the connection process of PANDORA-CO. Therefore, PANDORA-CO has more available information because of its connection aggregation process. That is also why PANDORA-CO achieves the best training efficiency.

The three different aggregators are then discussed. Fig.~\ref{fig:TSNE} shows the aggregators and their corresponding t-SNE (t-distributed Stochastic Neighbor Embedding) results of PANDORA-HA, PANDORA-SU, and PANDORA-CO. We list the aggregation process on the left side and corresponding results on the right side in each figure. Different colors represent different risk levels. Specifically, red refers to the high-risk level region, blue represents the medium-risk region, yellow stands for the low-risk region, and green refers to the risk-free region. Results shown in Fig.~\ref{fig:TSNE} indicate that the proposed methods can achieve significant features for each node. Different categories can be seen clearly in t-SNE results. It can also be seen that high-risk level areas are significantly clustered together, which verifies that PANDORA can well predict the risk levels of these regions.

\begin{table}[!htbp]
	\centering
	\caption{Results of ablation study. The hyperparameters are set to $\alpha$ =0.003, max\_epoch = 500, and ED = 64, respectively.}
	\resizebox{0.50\textwidth}{!}{
		\begin{tabular}{cccc}
			\toprule
			Methods & Accuracy& Iteration &Highest Accuracy\\
			\midrule
			GCN  & 80.50\%&300&80.83\% \\
			PANDORA-HA & 82.00\%&179& 82.50\%\\
			PANDORA-SU & \textbf{82.17\%} &127&82.17\%\\
			PANDORA-CO & 81.67\% &115&81.67\%\\
			w/o SFT & 76.00\% &223&76.33\%\\
			w/o AFT & 81.33\% &132&81.67\%\\
			\bottomrule
	\end{tabular}}
	\label{tab:ablation}
\end{table}

\begin{table}[!htbp]
	\centering
	\caption{Impacts of node attributes on model performance. The hyperparameters are set to $\alpha$ =0.003, max\_epoch = 500, and ED = 64, respectively.}
	\begin{tabular}{cccc}
		\toprule
		& Accuracy & Iteration & Highest Accuracy \\
		\midrule
		PANDORA-HA & \textbf{82.00\%} & 179 & 82.50\%  \\
		w/o geo-clim  & 80.67\% & 155 & 80.67\% \\
		w/o med & 81.00\% & 171 & 81.67\%  \\
		w/o eco & 81.17\% & 158 & 81.33\%  \\
		w/o demo & 80.83\% & 163 & 81.33\%  \\
		\midrule
		PANDORA-SU & \textbf{82.17\%} & 127 & 82.17\%  \\
		w/o geo-clim & 81.33\% & 118 & 81.50\% \\
		w/o med & 81.00\% & 124 & 81.50\%  \\
		w/o eco & 81.83\% & 133& 82.17\%  \\
		w/o demo & 81.83\% & 140 & 81.83\%  \\
		\midrule
		PANDORA-CO & \textbf{81.67\%} & 115 & 81.67\%  \\
		w/o geo-clim & 80.50\% & 104 & 80.67\% \\
		w/o med & 81.17\% & 129 & 81.67\%  \\
		w/o eco & 81.33\% & 116 & 81.83\%  \\
		w/o demo  & \textbf{81.67\%} & 118& 82.00\%  \\
		\bottomrule
	\end{tabular}%
	\label{tab:res_att_ablation}%
\end{table}%

We further conduct the ablation study to explore the effectiveness of the key strategies in our approach. Specifically, we use SFT and AFT separately to perform the prediction task to verify their validity. The results of ablation study are shown in Table.~\ref{tab:ablation}. Therein, $\alpha=0.003$, and max epoch is set as 500. It can be seen from the table that both variants perform less well than the three full models, which demonstrate the important role of the two feature tensors. 
The impacts of node attributes on model performance are then discussed. We remove one attribute from each of the four attributes. The experimental results can be seen in Table.~\ref{tab:res_att_ablation}, where $demo, med, geo-clim$ and $eco$ represent the four node attributes mentioned above. From the table, we can see that all four node attributes are useful, proving the effectiveness of the strategy of integrating node attributes. By further analysis we can find that medical conditions are more important among the four node attributes. It also shows that the medical condition of a region is a key factor in pandemic prediction.

\subsubsection{Dynamic Graphs}

We use Accuracy (Acc), Macro-Precision (Macro-Pre), Macro-Recall (Macro-Rec), and Macro-F1 as evaluation metrics to evaluate the performance of PANDORA on dynamic graphs. The experimental results are shown in Table~\ref{tab:res_in_dy}. Bolded results represent the outperformance method. It is shown that our proposed PANDORA has better performance than other baselines on dynamic graphs. Among them, PANDORA-HA achieves the best performance in Acc, Macro-Pre, and Macro-F1. PANDORA-CO performs best on Macro-Rec. Moreover, PANDORA achieves better results on dynamic graphs than static graphs. In terms of accuracy, PANDORA is generally higher than baselines (around 7\%-21\%), indicating that our method can better forecast high-risk level regions. As for Macro-Pre, Macro-Rec, and Macro-F1, PANDORA also performs better than baselines (around 9\%-23\%).
ST-GCN pays more attention to forecasting in time series scenarios and thus may not have a closer simulation for spatial (the spread of epidemics) as PANDORA. Graph WaveNet solves the long-term dependency problem. Since the period we choose is not that long (because the forecasting task requires short-term prediction), Graph WaveNet performs worse than PANDORA. AGCRN pays more attention to the pattern of each node (county), but the spread of the pandemic will not follow the same pattern in each county. On the contrary, the spread patterns of the pandemic formulated with network motifs are more suitable for this forecasting task.

\section{Conclusion}
\label{sec:con}

Predicting certain regions' infection risk is a great strategy in combatting the COVID-19 pandemic. We have proposed PANDORA to forecast the infection risk level of COVID-19 with comprehensive factors. By constructing a geographical network based on geographical positions and flight information, these factors are formulated with node attributes and structural features. We take four kinds of vital elements as node attributes: climate, medical condition, economic status, and human mobility. Network motif is explored to formulate geographical relationships among different regions and transportation frequency. We have also devised three different aggregators to better aggregate the node attribute feature tensor and structural feature tensor together. The three aggregators are the Hadamard aggregator, Summation aggregator, and Connection aggregator. Experimental results show that no matter which kind of aggregator is used, the proposed methods (PANDORA-HA, PANDORA-SU, and PANDORA-CO) consistently outperform the baseline method in terms of both accuracy and convergence speed.

\section*{Acknowledgment}
The authors would like to thank Chen Zhang and Mingliang Hou at the Dalian University of Technology for their help with experiments. 
\bibliographystyle{ieeetr}
\bibliography{ref}

\begin{IEEEbiography}[{\includegraphics[width=1in,height=1.25in]{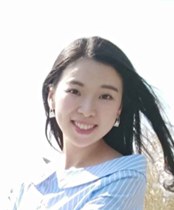}}]{Shuo Yu} (M'20) received the B.Sc. degree and M.Sc. degree in School of Science, Shenyang University of Technology, Shenyang, China. She is Associate professor in School of Computer Science and Technology, Dalian University of Technology, Dalian, China. Dr. Shuo Yu is currently working as a Postdoctoral Research Fellow in School of Software, Dalian University of Technology. She has published over 30 peer-reviewed papers in ACM/IEEE conferences and journals. Her research interests include network science, data science, computational social science, and knowledge graphs.
\end{IEEEbiography}

\begin{IEEEbiography}[{\includegraphics[width=1in,height=1.25in]{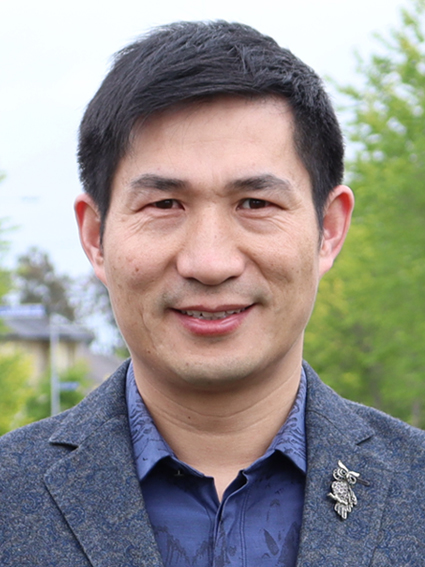}}]{Feng Xia} (M'07-SM'12) received the BSc and PhD degrees from Zhejiang University, Hangzhou, China. He was Full Professor and Associate Dean (Research) in School of Software, Dalian University of Technology, China. He is Associate Professor and former Discipline Leader (IT) in School of Engineering, IT and Physical Sciences, Federation University Australia. Dr. Xia has published 2 books and over 300 scientific papers in international journals and conferences (such as IEEE TAI, TKDE, TNNLS, TBD, TCSS, TNSE, TETCI, TC, TMC, TPDS, TETC, THMS, TVT, TITS, TASE, ACM TKDD, TIST, TWEB, TOMM, WWW, AAAI, SIGIR, CIKM, JCDL, EMNLP, and INFOCOM). His research interests include data science, artificial intelligence, graph learning, anomaly detection, and systems engineering. He is a Senior Member of IEEE and ACM. 
\end{IEEEbiography}

\begin{IEEEbiography}[{\includegraphics[width=1in,height=1.25in]{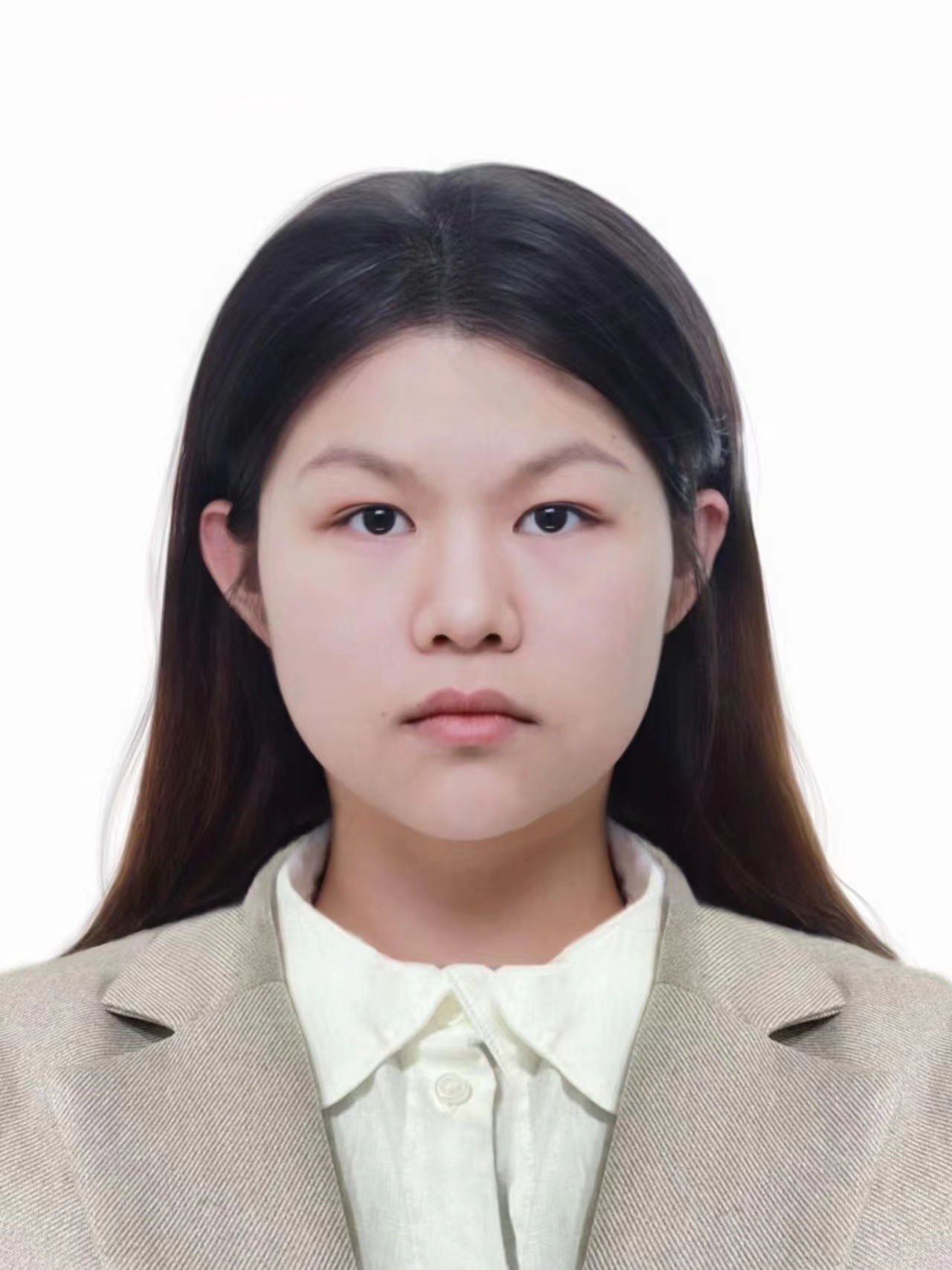}}]{Yueru Wang} is currently pursuing her Bachelor's degree in Department of Mathematics, National Tsing Hua University, Taiwan. Her research interests include computational intelligence and data science.
\end{IEEEbiography}

\begin{IEEEbiography}[{\includegraphics[width=1in,height=1.25in]{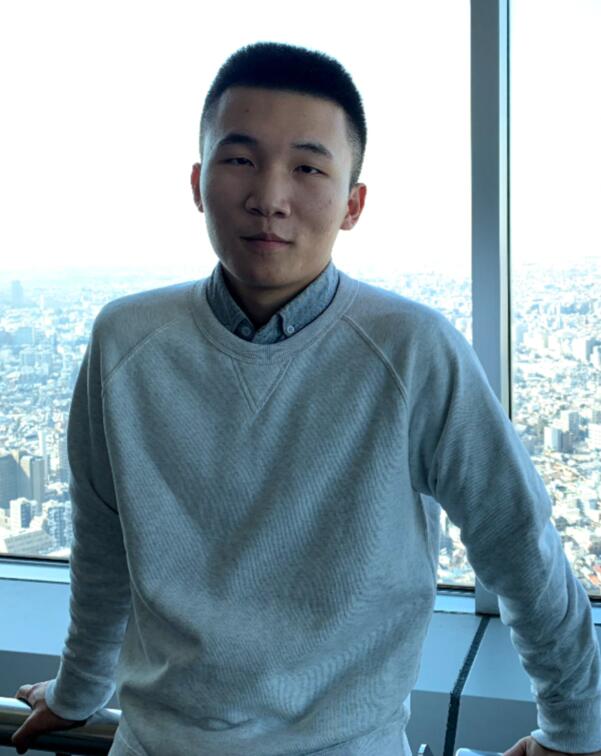}}]{Shihao Li} received the B.Sc. degree from Shenyang Normal University, Shenyang, China in 2019. He is currently pursuing the Master degree in Software Engineering at Dalian University of Technology, Dalian, China. His research interests include network science, knowledge graphs and data analytics.
\end{IEEEbiography}

\begin{IEEEbiography}[{\includegraphics[width=1in,height=1.25in]{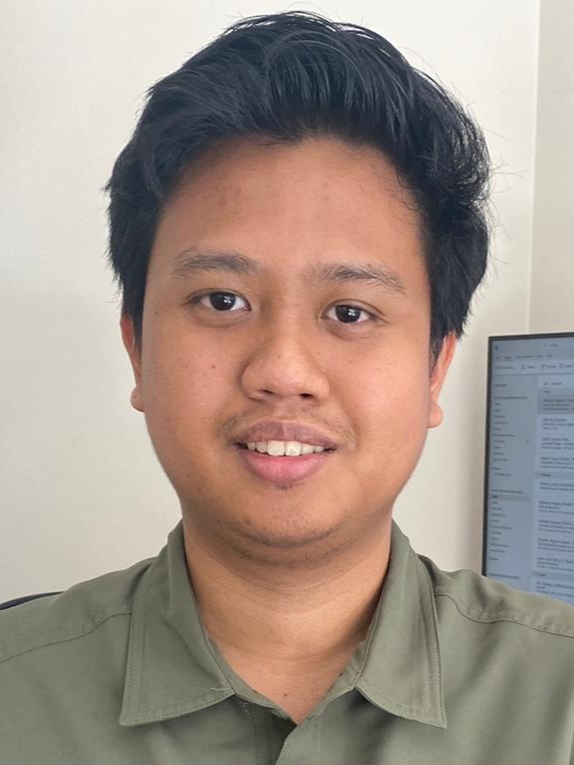}}]{Falih Gozi Febrinanto} received the Bachelor of Computer Science (S.Kom), from the University of Brawijaya, Malang, Indonesia, in 2018 and the Master of Technology (M.Tech) from Federation University Australia, Ballarat, Australia, in 2021. He is currently Ph.D. in Information Technology student at Federation University Australia and CSIRO’s Data61, Australia. His research interest includes Graph Learning, Lifelong Learning, Deep learning, and Human-centric Cyber Security.
\end{IEEEbiography}

\begin{IEEEbiography}[{\includegraphics[width=1in,height=1.25in]{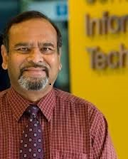}}]{Madhu Chetty} received both his Master degree and PhD from Nagpur University, India. He is currently an Associate Professor in School of Engineering, IT and Physical Sciences, Federation University Australia. He works extensively in the areas of graphical network models, computational intelligence, data analytics and their applications to areas such as bio- and health- informatics, and engineering systems. He has published over 160 scientific papers. He was Chair and Vice-Chair of international professional committees and currently serving as Vice Chair of IEEE Victorian Section in Australia.
\end{IEEEbiography}

\end{document}